\documentclass{PoS}
\usepackage{macros,epsfig}

\PoS{PoS(LAT2005)014}

\title{Twisted mass lattice QCD: Recent developments and results}

\ShortTitle{Twisted mass lattice QCD}

\author{\speaker{Andrea Shindler}\\
        NIC, DESY Platanenallee 6 \\
        15738 Zeuthen, Germany \\
        E-mail: \email{andrea.shindler@desy.de}}

\abstract{I review recent theoretical developments and numerical results of
          twisted mass QCD. I argue that, combined with an efficient
          algorithm, twisted mass QCD can be
          an attractive QCD lattice action, to perform large scale simulations
          at small pion masses, where a matching with chiral
          perturbation theory can be performed. Open issues like flavour
          breaking effects are also addressed.

\vspace{1.0cm}
{\large
\texttt{DESY 05-216}\\
\texttt{SFB/CPP-05-73}}
}

\FullConference{XXIIIrd International Symposium on Lattice Field Theory\\
		 25-30 July 2005\\
		 Trinity College, Dublin, Ireland}

\begin{document}

\section{Cutoff effects and renormalization}

Phenomenological results from simulations of lattice QCD to compare with
experiments should be obtained with all the systematic uncertainties 
under control.
The first requirement is to have an efficient algorithm to simulate 
$N_{\rm f}=2$ dynamical light quarks with the possibility to include 
1(+1) heavier quarks. The algorithm should allow, in a reasonable time, 
to reach small pion masses 
($m_\pi < 300$ MeV) where a matching with chiral perturbation theory ($\chi$PT)
should become possible, and to simulate a large enough volume ($L\ge2$ fm).
The second requirement is to have a lattice action with good scaling
and simplified renormalization properties, as close as possible to
the renormalization of continuum QCD.
The topic I want to address in this contribution is if
lattice twisted mass QCD (tmQCD), combined with a suitable algorithm, 
is a possible lattice action that fulfills these requirements.

\subsection{Lattice QCD action}
\label{sec:latticeqcd}

Despite the only rather recent interest, the tmQCD fermionic lattice action has
a long history. It was introduced in \cite{Aoki:1984qi} as a tool to study spontaneous
parity and flavour symmetry breaking. In \cite{Frezzotti:2000nk} it was proved
that lattice tmQCD is an alternative discretization of lattice QCD.
The lattice QCD action
\be
S = S_g[U] + S_F[U,\psi,\psibar]
\ee
has a fermionic part given by tmQCD
\be
S_F = a^4\sum_x \big\{\psibar(x)[D[U] + m_0 +i\mu\gamma_5\tau^3]\psi(x)\big\} 
\label{eq:tmQCD}
\ee
and for the moment we leave unspecified the gauge part $S_g$.
In eq. (\ref{eq:tmQCD}) $D[U]$ is the massless Wilson-Dirac operator
\be
D[U] = \frac{1}{2} [\gamma_\mu(\nabla_{\mu} + \nabla_{\mu}^*) - a
      \nabla_{\mu}^* \nabla_{\mu}]
\ee
$m_0$ is the untwisted bare quark mass parameter, $\mu$ is the bare twisted
quark mass, and $\tau^3$ is the third Pauli matrix acting in flavour space.
In \cite{Frezzotti:2001ea} it was shown that the standard framework of the Symanzik
improvement program, works in the similar way as for usual Wilson fermions. In
particular for spectral quantities no further improvement coefficients are
needed. A set of scaling tests have been performed, using the
non-perturbatively improved clover action with twisted mass, in small
\cite{DellaMorte:2001ys} and large \cite{DellaMorte:2001tu} volume, confirming
that the usual Symanzik improvement program can be applied also for tmQCD.

In a remarkable paper of Frezzotti and Rossi \cite{Frezzotti:2003ni} a step forward was made. 
It was proved that parity even correlators of multiplicatively renormalizable
fields, are free from O($a$) effects, and so no improvement coefficients are needed
(automatic O($a$) improvement), if
the target continuum theory is fully twisted \footnote{To obtain automatic
O($a$) improvement in \cite{Frezzotti:2003ni} also other possibilities were
exploited. Here we will concentrate on the automatic O($a$) improvement that
is used in numerical simulations.}.
The proof in \cite{Frezzotti:2003ni} is based on a set of spurionic symmetries
of the lattice action. Here we give a simpler proof based 
on the symmetries of the continuum QCD action (see appendix A of
\cite{Frezzotti:2005gi} for an analogous proof).
The Symanzik
\cite{Symanzik:1981hc,Symanzik:1983dc,Symanzik:1983gh,Sheikholeslami:1985ij,Luscher:1996sc} 
effective action reads
\be
S_{\rm eff} = S_0 + aS_1 + \ldots
\ee
and we are interested in a continuum target theory where the physical quark mass is fully given by
the renormalized twisted mass $\mu_{\rm R}$ (fully twisted theory)
\be
S_0 = \int d^4x \psibar(x) \big [ \gamma_\mu D_\mu + {\rm i} \mu_{\rm R}
\gamma_5 \tau^3 \big] \psi(x)
\label{eq:ctmQCD}
\ee
The correction terms in the effective action are given by
\be
S_1 = \int d^4y  {\mathcal L}_1(y) \qquad {\mathcal L}_1(y) = \sum_i c_i
  {\mathcal O}_i(y)
\ee
where the dimension five operators classified on the basis of the symmetries
of the lattice action are given by \cite{Frezzotti:2001ea}
\be
{\mathcal O}_1 =
  \psibar\sigma_{\mu\nu}F_{\mu\nu}\psi  \qquad  {\mathcal O}_2 =
  \mu^2\psibar \psi \qquad  {\mathcal O}_3 = \Lambda^2\psibar \psi
\label{eq:sym_op}
\ee
where $\Lambda$ is an energy scale of the order of the QCD scale $\Lambda_{\rm QCD}$.
The operator ${\mathcal O}_1$ is the usual clover term. The operators
${\mathcal O}_2$ and ${\mathcal O}_3$ are related to the renormalization of the untwisted
quark mass. 
Since we are interested in a continuum target theory where the untwisted quark
mass vanishes, the operator ${\mathcal O}_3$
parameterizes the mass independent O($a$) uncertainties in the critical mass.
We consider now a general multiplicatively renormalizable multilocal field
that in the effective theory is represented by the effective field
\be
\Phi_{\rm eff} = \Phi_0 + a \Phi_1 + \ldots
\ee
A lattice correlation function of the field $\Phi$ to order $a$ is given by
\be
\langle \Phi \rangle = \langle \Phi_0 \rangle_0 + a \int d^4y \langle \Phi_0
     {\mathcal L}_1(y) \rangle_0 + a \langle \Phi_1 \rangle_0 + \ldots
\label{eq:sym_exp}
\ee
where the expectation values on the r.h.s are to be taken in the continuum
theory with action $S_0$.
The key point is that the continuum action (\ref{eq:ctmQCD}) is symmetric
under the following parity transformation
\be
\psi(x) \longrightarrow \gamma_0(i \gamma_5\tau^3)\psi(x_0,-{\bf x})
\label{eq:parity}
\ee
\be
\psibar(x) \longrightarrow \psibar(x_0,-{\bf x})(i\gamma_5\tau^3)\gamma_0
\label{eq:paritybar}
\ee
and that all the operators in eq. (\ref{eq:sym_op}), of
the Symanzik expansion of the lattice action, are odd under the parity
symmetry of the continuum action. 
If the operator $\Phi_0$ is parity even, the second term in the r.h.s. of
eq. (\ref{eq:sym_exp}) vanishes, and $\Phi_1$, being of one dimension higher, 
is parity odd: for the same reason the third term in the r.h.s of
eq. (\ref{eq:sym_exp}) vanishes. Possible contact terms
coming from the second term amount to a redefinition of $\Phi_1$ and so do not
harm the proof.

It is then also clear that in order to achieve automatic O($a$) improvement, the
continuum target theory must have a vanishing untwisted quark mass $m_{\rm
  R}$, otherwise the standard mass term $m_{\rm R} \psibar \psi$ 
will break the parity symmetry of the continuum action defined before. The
most natural way to achieve this on the lattice is
by setting the untwisted bare quark mass to
its critical value $m_0 = m_{\rm c}$.
The proof also shows that a possible uncertainty of O($a$) in the critical
mass does not wash out automatic O($a$) improvement since these uncertainties,
are odd under parity.
A remark is in order now. 
We take the polar mass defined in \cite{Frezzotti:2001ea}
\be
M = \sqrt{\mu^2 + m_{\rm q}^2} = \sqrt{\mu^2 + (\eta_1 a \Lambda^2)^2}; \qquad m_{\rm q} = m_0 - m_{\rm c}
\ee
where the $\eta_1$ term parameterizes the mass independent O($a$) uncertainties in
the value of the untwisted quark mass $m_{\rm q}$.
Expanding in powers of $a$ we have
\be
M \simeq \mu\Big[1+\frac{\eta_1 a^2\Lambda^4}{2\mu^2} +
  O(a^4)\Big]
\label{eq:pole_exp}
\ee
We observe immediately that as soon $\mu < a\Lambda^2$, even if parametrically
O($a$) terms are absent in (\ref{eq:pole_exp}), there is a term of O($a^2$) with a
coefficient that tends to diverge as soon $\mu$ is made smaller and smaller.
From this example we can conclude that to have an effective automatic O($a$) 
improvement, without big O($a^2$) effects, with a generic choice of the critical mass, such that the
uncertainties in the untwisted quark mass are of order $a\Lambda^2$, we need to have the
constraint $\mu > a\Lambda^2$.
It has been shown in \cite{Frezzotti:2005gi} that these cutoff effects that
diverges at small quark masses, so called infrared divergent (IR) cutoff
effects, are a general property of tmQCD. These dangerous cutoff effects are
removed by an appropriate choice of the critical mass.

\subsection{O($a$) improvement and small pion masses}
\label{sub:1.2}

The O($a$) uncertainties of the untwisted quark mass depend on how the
critical line is fixed, hence the choice of the critical mass has to be
discussed with care.
The issue was raised by the work of Aoki and B\"ar \cite{Aoki:2004ta} and by
the numerical results obtained in \cite{Bietenholz:2004wv}. This problem has
been further analyzed in several aspects  
\cite{Sharpe:2004ny,Frezzotti:2005gi,Sharpe:2005rq}. In
\cite{Aoki:2004ta,Sharpe:2004ny,Sharpe:2005rq} the theoretical framework is twisted mass
chiral perturbation theory (tm$\chi$PT) \cite{Munster:2003ba} where the cutoff effects are included
in the chiral lagrangian along the lines of
\cite{Sharpe:1998xm,Rupak:2002sm}. In this framework a power counting
scheme that includes quark mass and lattice spacing has to be specified. In
particular in \cite{Aoki:2004ta} the power counting was $\mu \sim a^2
\Lambda^3$ while in \cite{Sharpe:2004ny} it was $\mu \sim a \Lambda^2$. 
We stress here that this approach for the description of lattice data, does
not require a continuum extrapolation, hence the power counting scheme does
not mean that $\mu$ goes to zero in the continuum limit but represents only an
order of magnitude equality.
Both these works \cite{Aoki:2004ta,Sharpe:2004ny}
agree on the fact that choosing the critical mass imposing a vanishing PCAC
quark mass
\be
m_{\rm PCAC} = \frac{\sum_{\bf x} \langle \partial_0 A_0^a(x) P^a(0)\rangle}
{2\sum_{\bf x} \langle  P^a(x) P^a(0)\rangle} \qquad a=1,2
\ee
where
\be
A_\mu^a(x) = \psibar(x)\gamma_\mu \gamma_5 {\tau^a \over 2} \psi(x)
\ee
\be
P^a(x) = \psibar(x)\gamma_5 {\tau^a \over 2} \psi(x)
\ee
allows to have automatic O($a$) improvement, and in particular down to quark masses that fulfill
$\mu \simeq a^2\Lambda^3$ for 
\cite{Aoki:2004ta} and $ a^2\Lambda^3 < \mu < a
\Lambda^2$ for \cite{Sharpe:2004ny}\footnote{We will see in section \ref{sec:phase} that
  the phase structure of $N_{\rm f} = 2$ dynamical Wilson fermions 
does not allow anyway the twisted mass to be smaller then $\mu_{\rm c} \sim
a^2\Lambda^3$.}.
In \cite{Frezzotti:2005gi} a Symanzik expansion along the lines of
\cite{Luscher:1996sc} was performed confirming the results of \cite{Aoki:2004ta,Sharpe:2004ny}.

\begin{figure}[htb]
\begin{center}
\epsfig{file=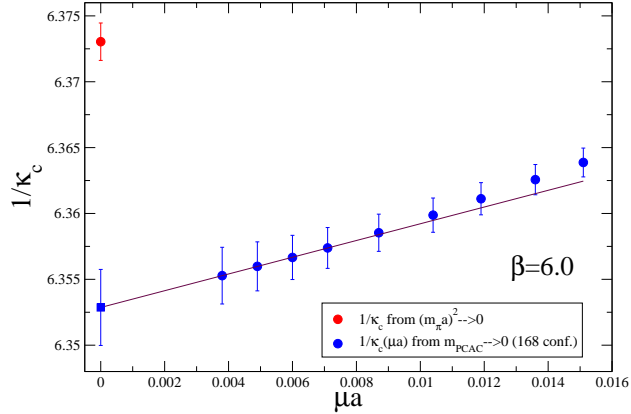,angle=270,width=0.6\linewidth}
\caption{Determination of the critical mass $m_{\rm c}$ ($\kappa_{\rm c}^{-1} =
2am_{\rm c} +8$) for given values of $\mu$ at $\beta=6.0$, and extrapolation to
$\mu=0$. The red point is the critical mass determined with method {\bf C}
(see text). The difference between the two determination of the critical mass
should be an O($a$).}
\label{fig:kappa_c}
\end{center}
\end{figure}

A possible practical procedure is then to compute for a fixed value of $\mu$ the
critical mass $m_{\rm c}$ from the vanishing PCAC mass,
and then to extrapolate the set of critical masses
obtained for different values of $\mu$ to $\mu = 0$ (method {\bf A}). 
This procedure has been used in \cite{Jansen:2005gf,Jansen:2005kk}. 
In fig. \ref{fig:kappa_c} a typical
extrapolation of the critical mass to $\mu=0$ is shown.
With this procedure the O($a$) uncertainties of the critical mass are fixed in
such a way that, for a generic value of $\mu$, the dangerous $a\Lambda^2$
cutoff effects in the untwisted quark mass are absent. 
The slope of the curve is proportional, as it has been recently discussed in
\cite{Aoki:2005ii,Sharpe:2005rq}, to O($a$) cutoff effects related to the
discretization errors of the PCAC mass.
We remind that it is not surprising that the PCAC mass is not
automatically O($a$) improved since it is an odd quantity under the parity
transformation of eqs. (\ref{eq:parity}, \ref{eq:paritybar}).

In \cite{Abdel-Rehim:2005gz} the extrapolation to
$\mu=0$ is not performed and each value of the critical mass has been used for
the corresponding value of $\mu$ used in the simulations (method {\bf B}).
With this method the O($a$) cutoff effects of the critical mass are obviously fixed in such a
way that the untwisted quark mass is always vanishing for all the simulation
points.

These two methods, even if they give
different cutoff effects to the critical mass, are perfectly good in order to
achieve automatic O($a$) improvement. 
Another possible way to fix the critical mass, expecially practical for
expensive dynamical simulations, is to compute the critical mass, using the
PCAC relation at the smallest value of $\mu$, and then use this critical mass
for all the simulation points at heavier masses.

Using methods {\bf A} and {\bf B} a set of quenched studies
\cite{Jansen:2003ir,Bietenholz:2004wv,Abdel-Rehim:2004gx,Abdel-Rehim:2005gz,Jansen:2005gf,Jansen:2005kk,Abdel-Rehim:2005qv} 
have been performed to check the result of \cite{Frezzotti:2003ni} 
and to gain experience with this formulation of lattice QCD.
\begin{figure}[htb]
\epsfig{file=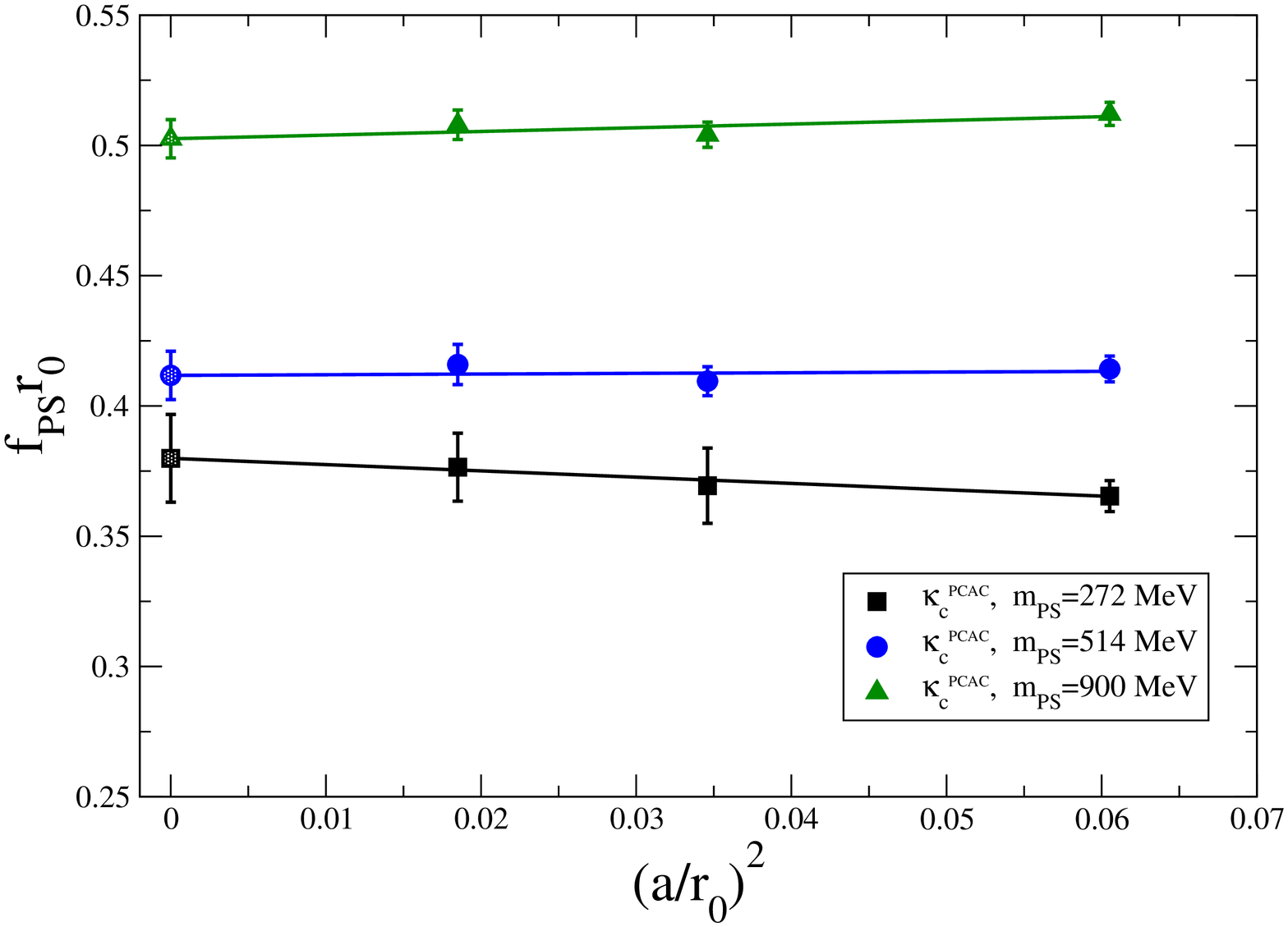,angle=270,width=0.5\linewidth}
\epsfig{file=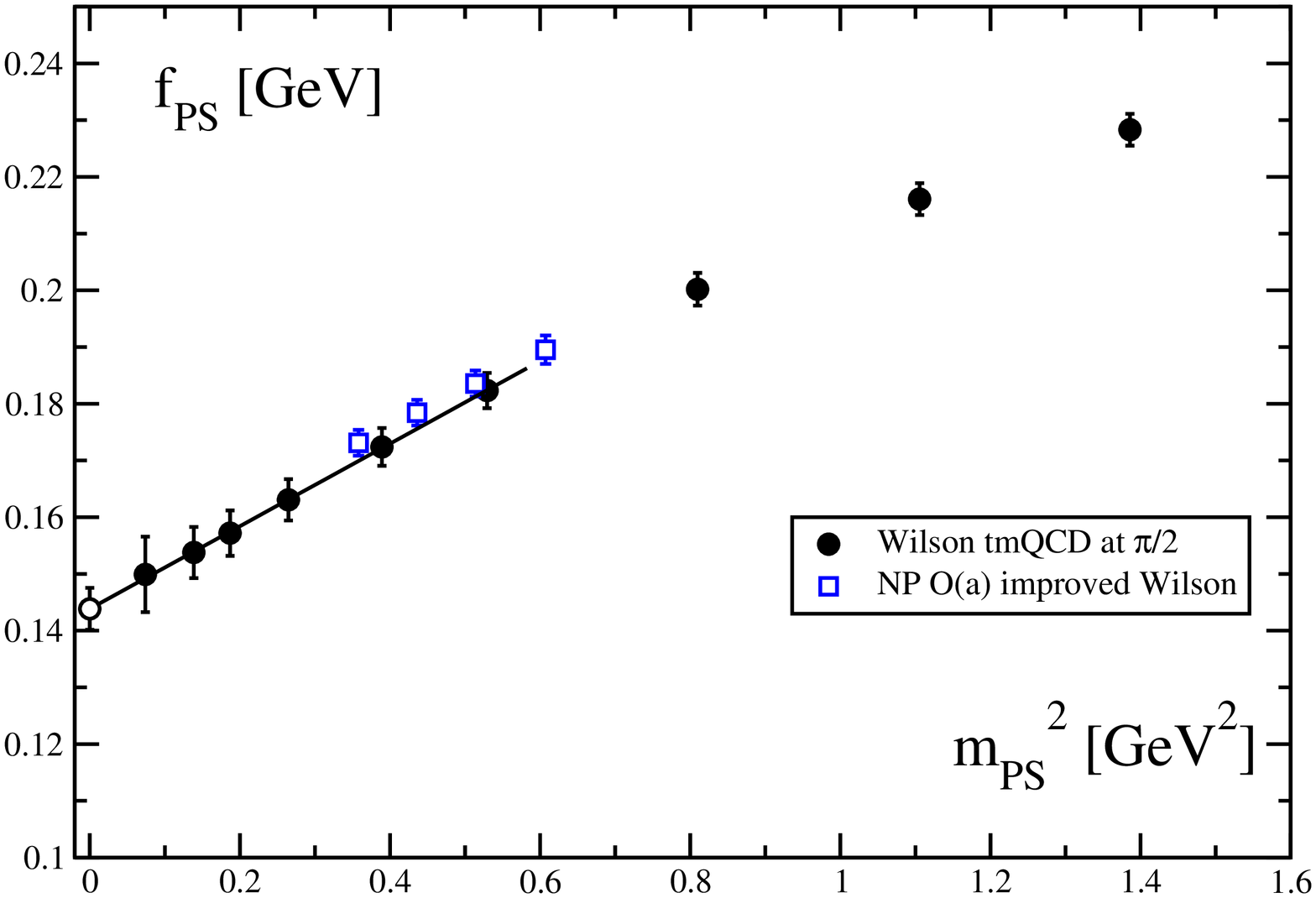,angle=270,width=0.5\linewidth}
\caption{Left panel: scaling behaviour of $r_0f_{\rm PS}$ for 3 fixed values
  of $r_0m_{\rm PS}$. Right panel: Continuum limit values for $f_{\rm PS}$
  as a function of $m_{\rm PS}^2$ in physical units. The empty squares are taken from \cite{Garden:1999fg}.}
\label{fig:fps}
\end{figure}
An interesting quantity to compute with tmQCD is the pseudoscalar decay
constant $f_{\rm PS}$. As it was noted in
\cite{DellaMorte:2001tu,Frezzotti:2001du,Jansen:2003ir}, 
the computation of $f_{\rm PS}$ does not require any renormalization constant,
in contrast of ordinary Wilson fermions, and moreover given automatic O($a$)
improvement, does not need the computation of any improvement coefficient. 
Thus the situation for this quantity is like with overlap fermions. 
In fig. \ref{fig:fps} (left panel) the
continuum limit of $r_0 f_{\rm PS}$ \footnote{The values of $r_0/a$, $r_0=0.5$
  fm being the Sommer scale \cite{Sommer:1993ce}, are taken from
  \cite{Guagnelli:1998ud}.}, the critical mass being computed with method {\bf A}, 
is shown as a function of $(a/r_0)^2$. The scaling is consistent with being of
O($a^2$), and moreover the O($a^2$) effects are rather small for all the pseudoscalar
masses investigated down to $m_{\rm PS} = 272$ MeV. The right panel of
fig. \ref{fig:fps} shows the chiral behaviour of the continuum pseudoscalar
decay constant, compared with the non-perturbatively O($a$) improved
data of \cite{Garden:1999fg}. We remark that this comparison is purely
illustrative since it is in the quenched approximation, and the simulations
with clover fermions had to stop around $m_{\rm PS} \simeq 500$ MeV due to the
appearance of exceptional configuration.
\begin{figure}[htb]
\begin{center}
\epsfig{file=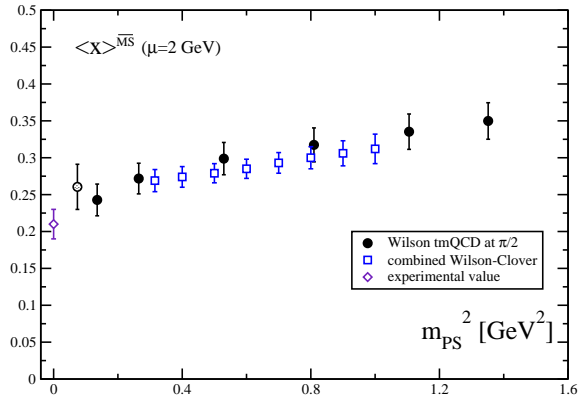,angle=270,width=0.6\linewidth}
\caption{$\langle x \rangle^{\msbar}(\mu= \ 2 \ GeV)$ extrapolated to the continuum as a function of
the pion mass. Open squares represent results that are obtained from a 
combined continuum
extrapolation of earlier Wilson and clover-Wilson simulations
\cite{Guagnelli:2004ga}. The filled circles represent results using
Wilson twisted mass fermions \cite{Capitani:2005aa}. The open circle denotes 
a result which is not corrected for finite size effects and the diamond corresponds 
to the experimental point.}
\label{fig:xpion}
\end{center}
\end{figure}
To see the potential of tmQCD another interesting phenomenological quantity
is the average momentum carried by valence quarks in a pion ($\langle
x\rangle$).
In \cite{Capitani:2005aa} results using tmQCD were presented. Here we concentrate on the
chiral behaviour in the continuum, having in mind that the renormalization has
been performed already in a non-perturbative way
\cite{Guagnelli:2003hw,Guagnelli:2004ga}. 
Fig. \ref{fig:xpion}
shows that in principle also for this quantity small pseudoscalar masses
$m_{\rm PS} < 300$ MeV can be reached, opening the possibility of a safe
chiral extrapolation.

\begin{figure}[htb]
\epsfig{file=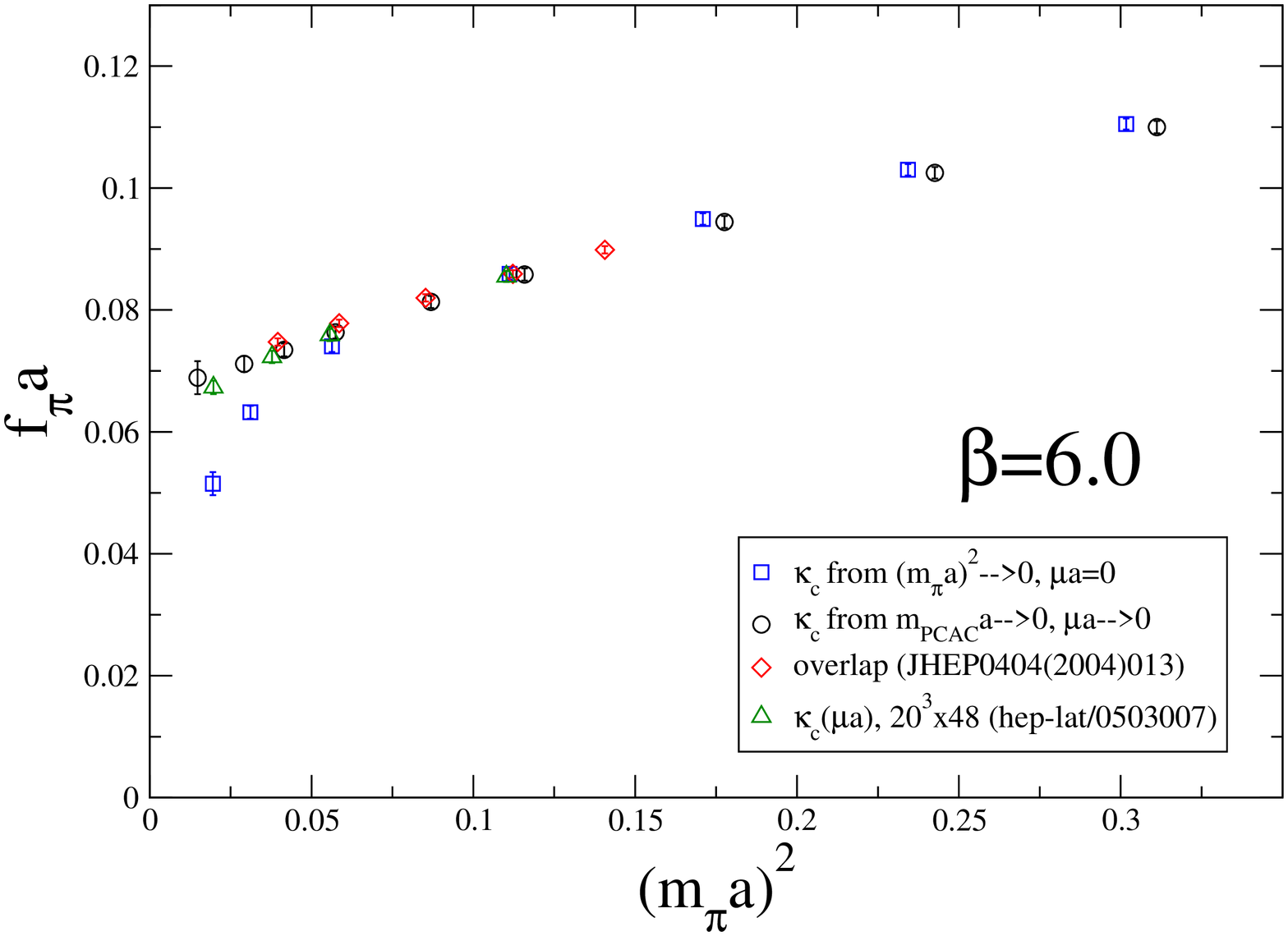,angle=270,width=0.5\linewidth}
\epsfig{file=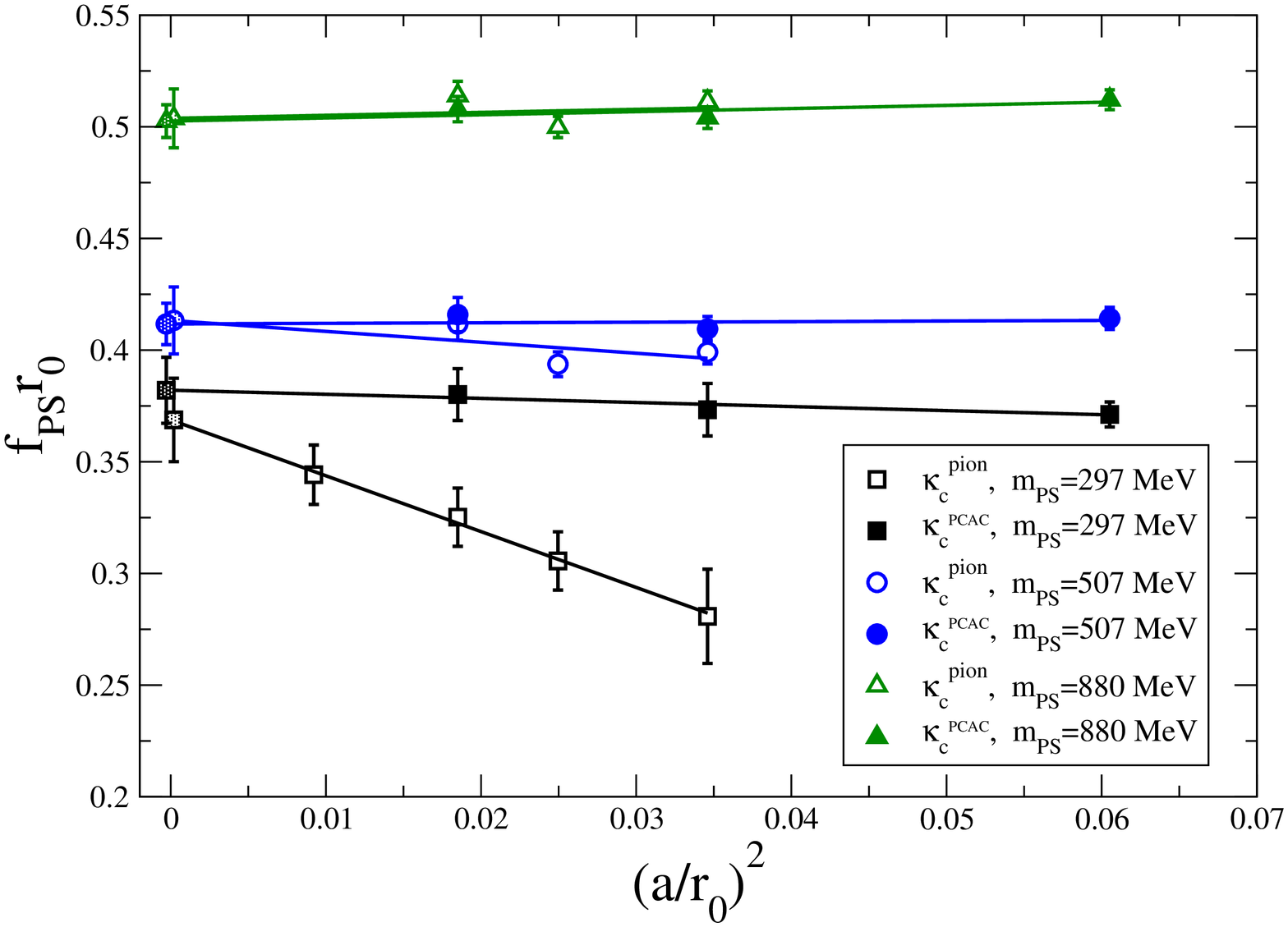,angle=270,width=0.5\linewidth}
\caption{Left panel: comparison of the chiral
behaviour at fixed lattice spacing ($\beta = 6.0$) of the pseudoscalar decay
constant computed using method {\bf A}, {\bf B}, {\bf C} and with
results obtained with overlap fermions. Right panel: unconstrained continuum
limit, for several values of fixed charge pion masses,
of $r_0f_{\rm PS}$ performed using method {\bf A} and {\bf C} to determine the
critical mass.}
\label{fig:bending}
\end{figure}
In \cite{Bietenholz:2004wv} to obtain automatic O($a$) improvement the critical
mass $m_{\rm c}$ was computed extrapolating the squared pseudoscalar mass to
the chiral limit using data from the pure Wilson theory (method {\bf C}). Using this
determination of $m_{\rm c}$ several quantities were computed. In particular in the
left panel of fig. \ref{fig:bending} there is a comparison of the chiral
behaviour at fixed lattice spacing ($\beta = 6.0$) of the pseudoscalar decay
constant computed using method {\bf A}, {\bf B}, {\bf C} and with
results obtained with overlap fermions \cite{Giusti:2004yp}. While methods
{\bf A}, {\bf B} and the overlap data are all consistent within the
statistical errors,\footnote{We recall here that since the comparison is made
  at fixed lattice spacing the data in principle could disagree due to
  different cutoff effects.} the data obtained using method {\bf C} to fix the
critical mass, show a ``bending'' towards the chiral limit.
The same phenomenon was observed also for the vector mass
\cite{Bietenholz:2004wv}.
The ``bending'' phenomenon appeared exactly when $\mu \simeq a
\Lambda^2$. Having in mind the caveat observed before in the proof of automatic O($a$)
improvement, this indicates that the extraction of the critical mass with
method {\bf C} leaves the dangerous $a\Lambda^2$ in the untwisted quark mass
uncanceled. This is numerically confirmed by the results of
\cite{Jansen:2005kk}, showed in the right panel of fig. \ref{fig:bending},
since using method {\bf A} and {\bf C} to determine the critical mass, a
consistent continuum limit is obtained, showing also that method {\bf C}
induces big O($a^2$) effects and a reduced scaling window.

\begin{figure}[htb]
\begin{center}
\epsfig{file=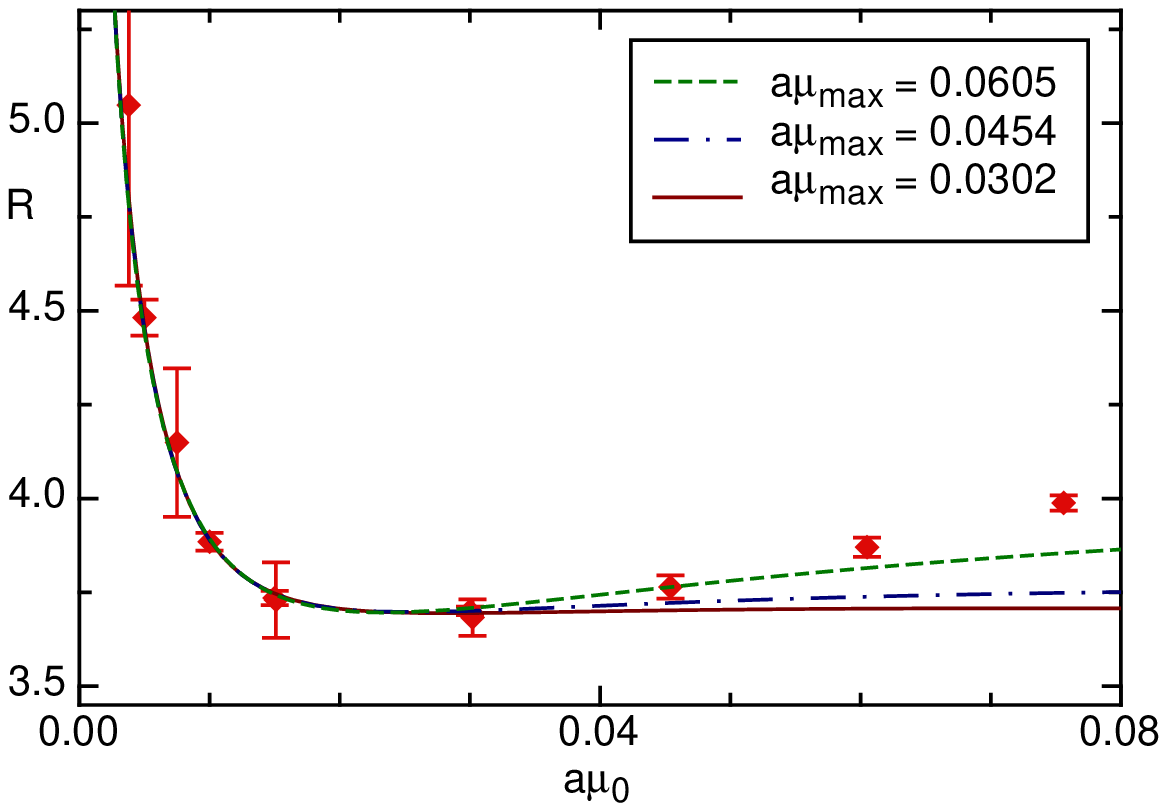,angle=0,width=0.42\linewidth} \hspace{0.8cm}
\epsfig{file=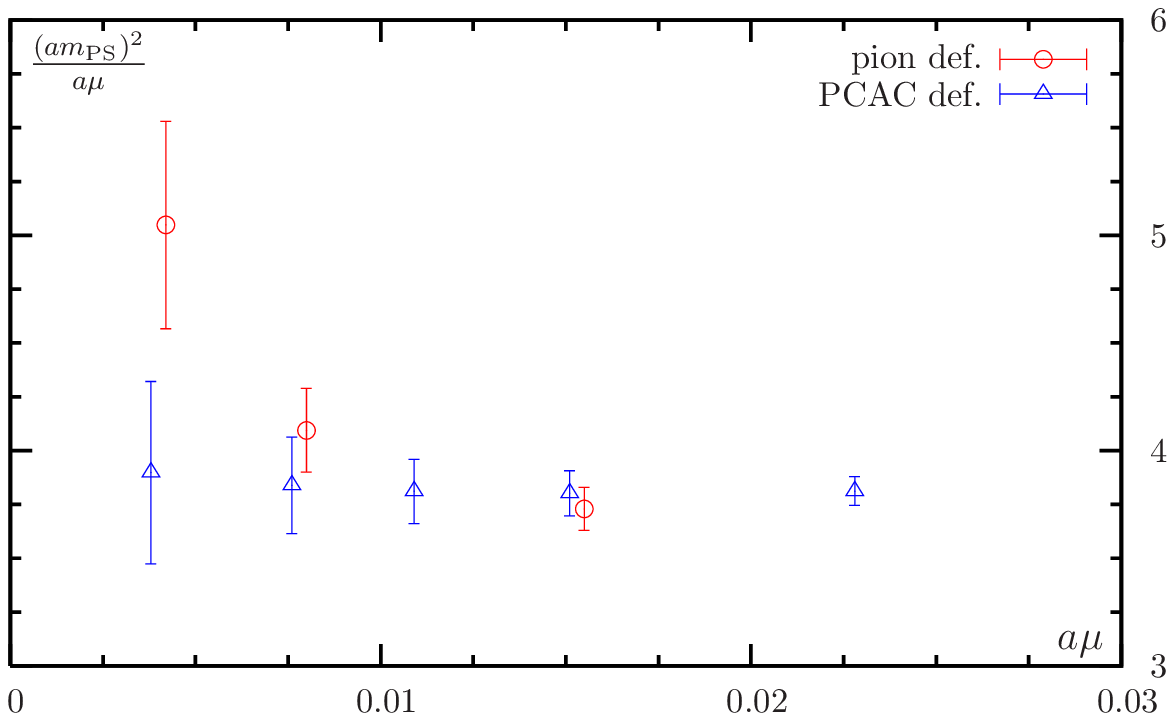,angle=0,width=0.45\linewidth}
\caption{Left panel: bending phenomenon at $\beta$=6.0 on the ratio $R={a^2
    m_{\rm PS}^2 \over a\mu}$ and its description with $\chi$PT. Right panel:
    comparison of the ration $R$ using method {\bf A} and {\bf C} to determine
    the critical mass.}
\label{fig:bar}
\end{center}
\end{figure}

\begin{figure}[htb]
\epsfig{file=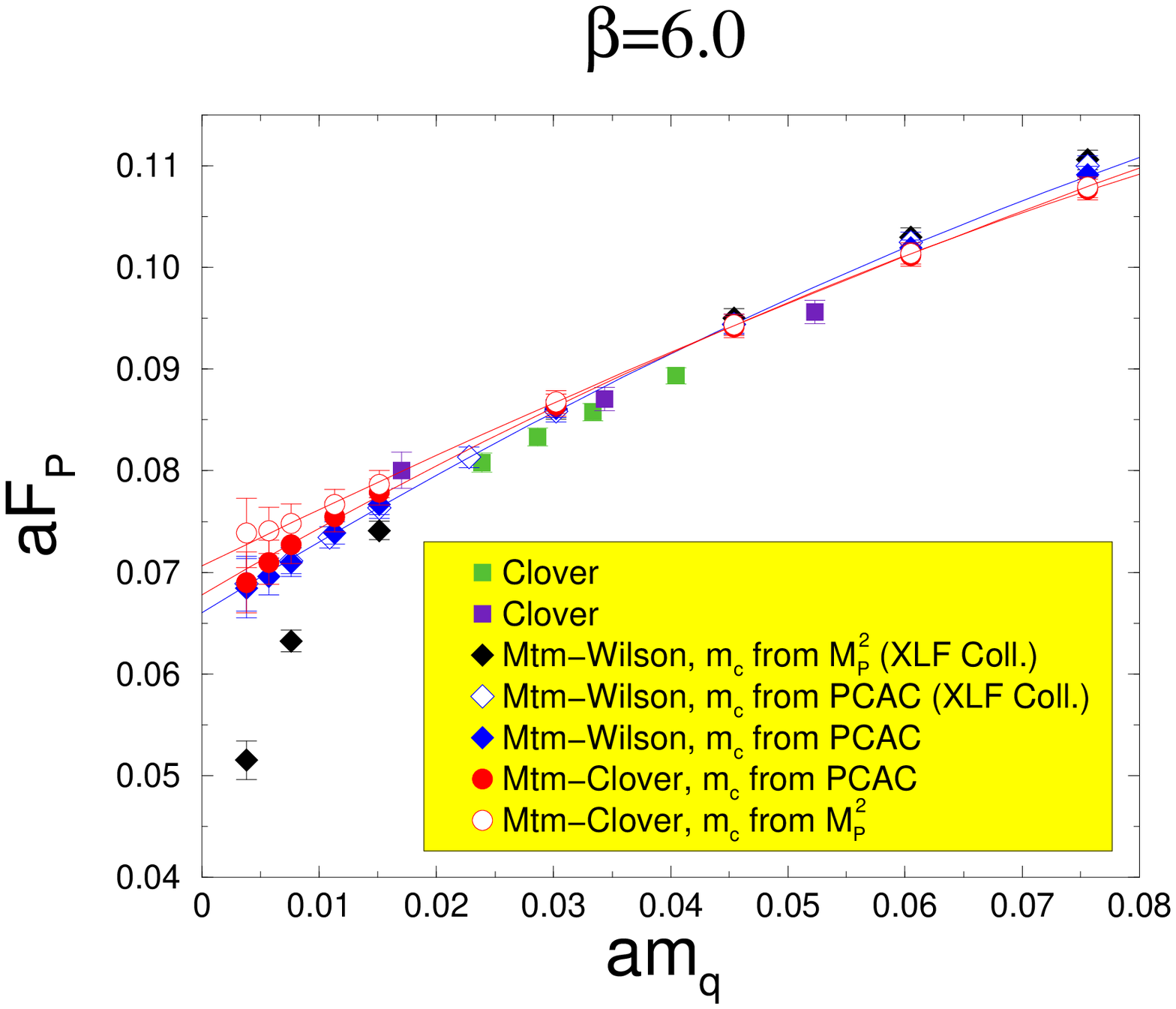,angle=0,width=0.5\linewidth}
\epsfig{file=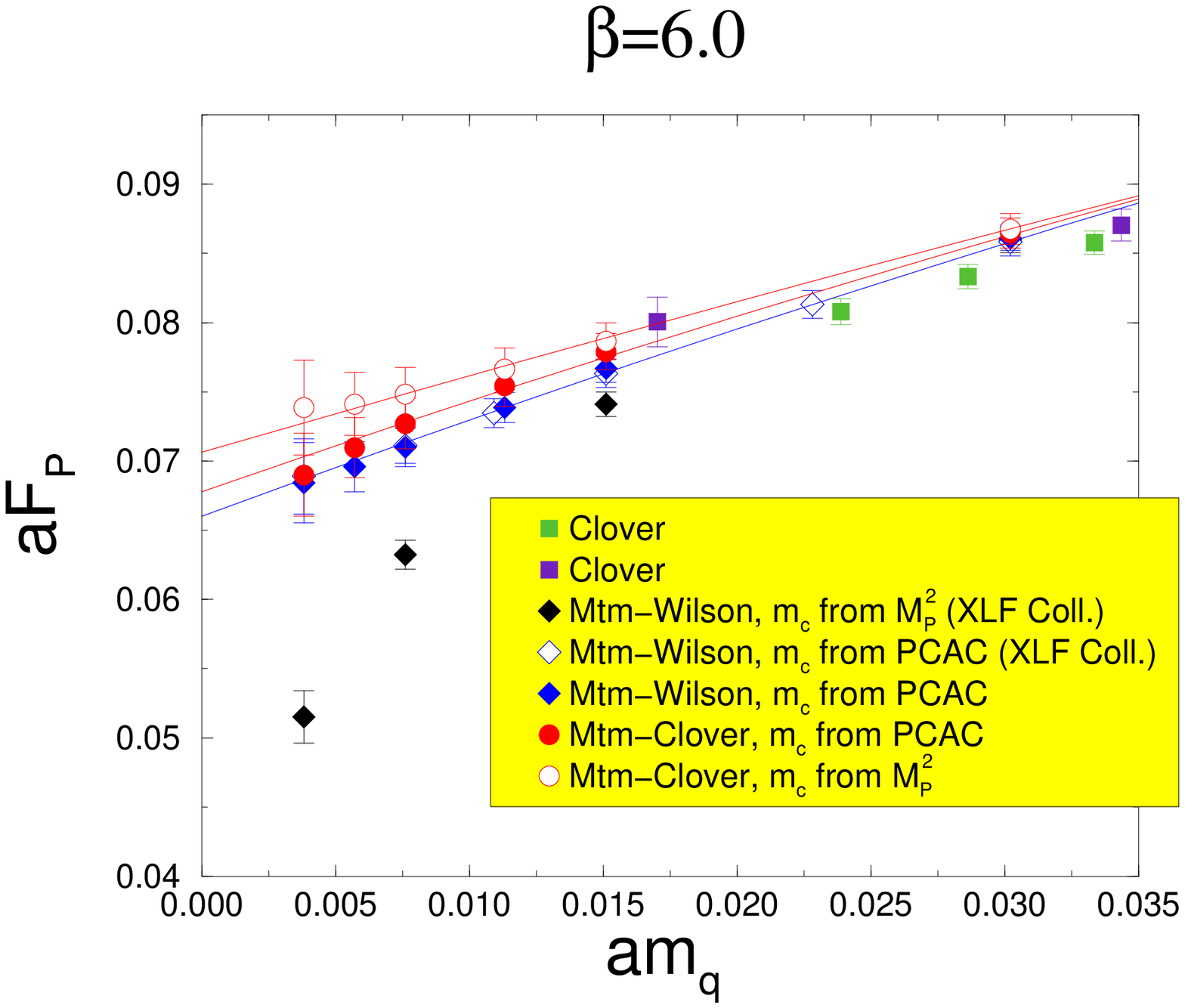,angle=0,width=0.47\linewidth}
\caption{Comparison of the chiral
behaviour at fixed lattice spacing ($\beta = 6.0$) of the pseudoscalar decay
constant computed using method {\bf A} and {\bf C} for both tmQCD and
non-perturbatively improved tmQCD.}
\label{fig:clover}
\end{figure}

A description of the ``bending'' phenomenon at fixed lattice spacing, 
has been obtained in \cite{Aoki:2005ii} using $\chi$PT, as it is shown in the
left panel of fig. \ref{fig:bar}, where a fit to available quenched data is
performed on the ratio $R={a^2 m_{\rm PS}^2 \over a\mu}$. 
This analysis shows also that $\chi$PT
theory is able to describe the lattice data up to $\mu \simeq 80$ MeV. It is
reassuring that using method {\bf A} to determine the critical mass and
restricting the data to the region were $\chi$PT is applicable the ratio $R$
is flat (right panel of fig. \ref{fig:bar}) 
consistently with continuum $\chi$PT (up to chiral logs).
In \cite{Frezzotti:2005gi}, based on the observation that the big
O($a^2$) effects come from uncanceled O($a$) of the PCAC mass, 
to eliminate the ``bending'' phenomenon has been proposed
to use a non-perturbatively improved tmQCD action. This approach has been
numerically tested in \cite{Lubicz}, and as it can be seen in
fig. \ref{fig:clover} indeed it confirms that the ``bending'' phenomenon also
in this case it is not present.

\subsection{Renormalization}

In \cite{Sint} it has been given a construction of a
Schr\"odinger functional (SF) with twisted boundary conditions that preserves the
nice properties of O($a$) improvement without bulk improvement coefficients;
(see \cite{Frezzotti:2005zm} for a possible alternative to this construction).
The construction is based on the consideration that in a finite volume with
suitable boundary condition the
Wilson theory in the chiral limit is O($a$) improved, and it makes use of orbifolding techniques
(see \cite{Taniguchi:2004gf} for an application of orbifolding techniques to Ginsparg-Wilson
fermions).

A simple way to visualize the construction is to repeat the proof of automatic 
O($a$) improvement given in section \ref{sec:latticeqcd}, 
where now since we are in a finite volume with suitable boundary conditions,
the twisted mass could be safely sent to zero.
Then the new boundary projectors \cite{Sint} $Q_{\pm} = {1 \over 2} (1+{\rm
i}\gamma_0 \gamma_5 \tau^3$) commute with the
previous parity transformation (as the twisted mass term in infinite volume).
It is very important to note that the new boundary projectors
can be obtained performing a chiral rotation of
the original projectors in the standard SF framework \cite{Sint:1993un}.
An important consequence is that
the running of the coupling constant, should be
identical to the running computed with the ``old'' SF \cite{DellaMorte:2004bc}.
The O($a$) uncertainties 
in the critical mass do not harm the O($a$) improvement.

\section{Flavour symmetry}

When tmQCD is used to define the standard QCD correlation functions some of the
physical symmetries are restored only in the continuum limit. In particular
flavour and parity symmetries.
The explicit breaking of flavour symmetry generates for example splitting between charged and neutral
pions, while the absence of parity symmetry, gives as a consequence the
appearance of states of opposite parity in the spectral decomposition of usual correlators.
Both these phenomena are expected to vanish, at maximal twist, with a
rate of O($a^2$) \cite{Frezzotti:2004wz}.
Here we concentrate on the flavour symmetry breaking.

To fix the notation we recall some basic definitions. The charged pseudoscalar
currents are given by 
\be
P^\pm(x) =\psibar(x)\gamma_5{\tau^\pm \over 2}\psi(x) \qquad  \tau^\pm = {\tau^1
  \pm {\rm i} \tau^2 \over 2}
\ee
and a possible interpolating field for the neutral pion is the scalar current
\be
S^0(x) = \psibar(x)\psi(x) .
\ee
The charged and neutral pseudoscalar masses can be extracted by the following correlators
\be
C_{\pi^{+}} (x_0) = a^3\sum_{\mathbf x} \langle P^+(x) P^-(0) \rangle \qquad
C_{\pi^{0}} (x_0) = a^3\sum_{\mathbf x} \langle S^0(x)S^0(0) \rangle 
\ee
\be
C_{\pi^{0}} (x_0) = a^3\sum_{\mathbf x} \big\{ \langle - \tr \big[ G(0,x) G(x,0)
  \big] + \tr \big[G(x,x)\big] \tr \big[G(0,0)\big] \rangle\big\}
\label{eq:pi0}
\ee
where $G(x,y)$ is the fermionic propagator.
In \cite{Jansen:2005cg} a pilot quenched study has been preformed to study
flavour breaking effects with tmQCD.
For the neutral pseudoscalar correlator in eq. (\ref{eq:pi0})
a first possibility is to study only
the connected part. In the quenched approximation it is still possible to
interprete the connected part in terms of local operators. The reason is that
one could think the connected part as coming from Wick contractions obtained
using the Osterwalder-Seiler (OS) \cite{Osterwalder:1977pc} action
\be
S_{\rm OS} = a^4\sum_x \big\{\psibar(x)[D[U] + m_0 +i\mu\gamma_5]\psi(x)\big\} .
\ee
This action has a trivial flavour structure and so does not present any
flavour breaking, and in particular the disconnected part of eq. (\ref{eq:pi0})
vanishes. We remark that this is not the neutral pseudoscalar meson of tmQCD,
but it is an interesting quantity to study with precise data on its own, in view
of a possible use of mixed actions (the OS action for the valence quarks and
tmQCD for the sea quarks).

\begin{figure}[htb]
\begin{center}
\epsfig{file=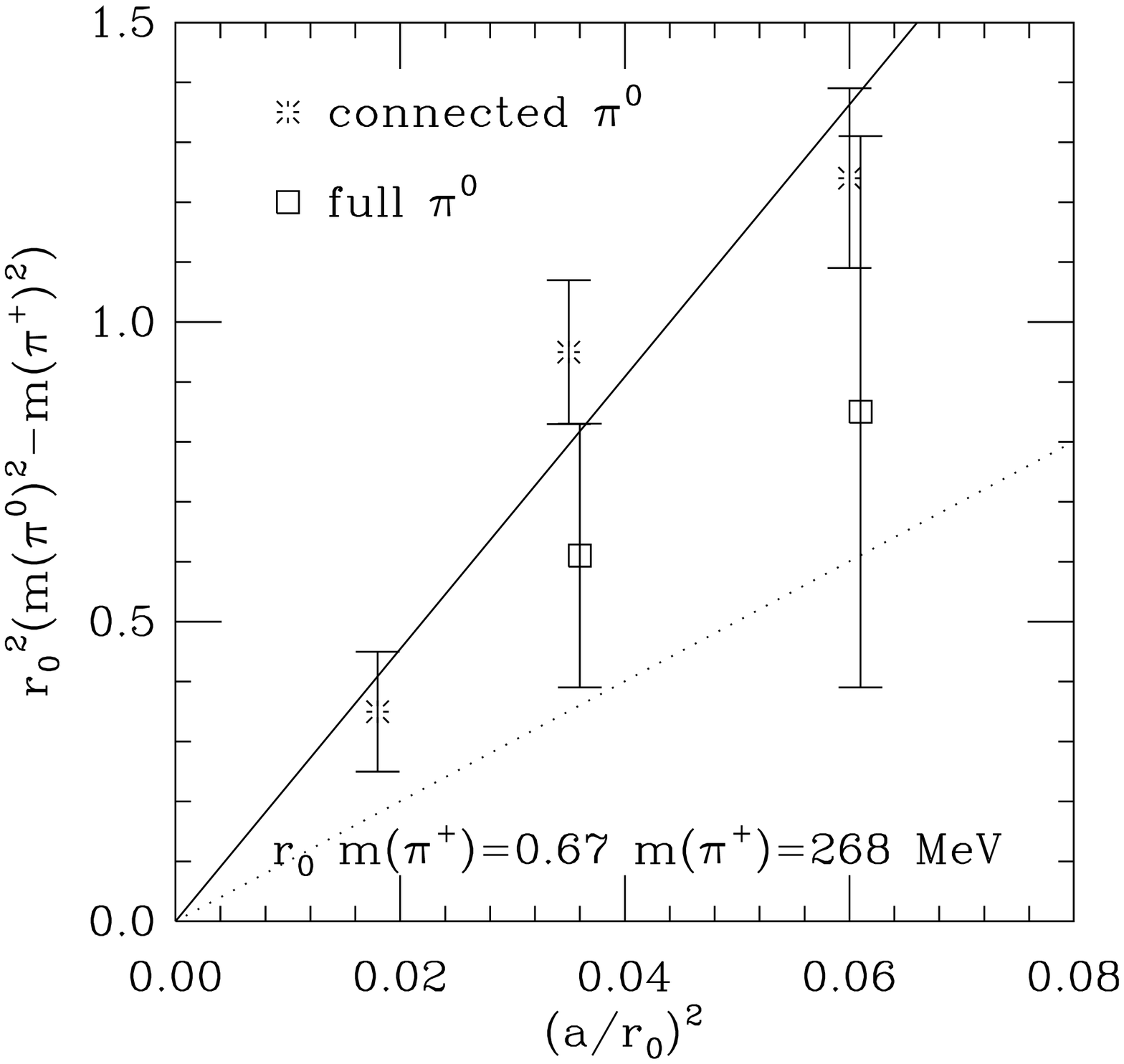,angle=0,width=0.48\linewidth}
\epsfig{file=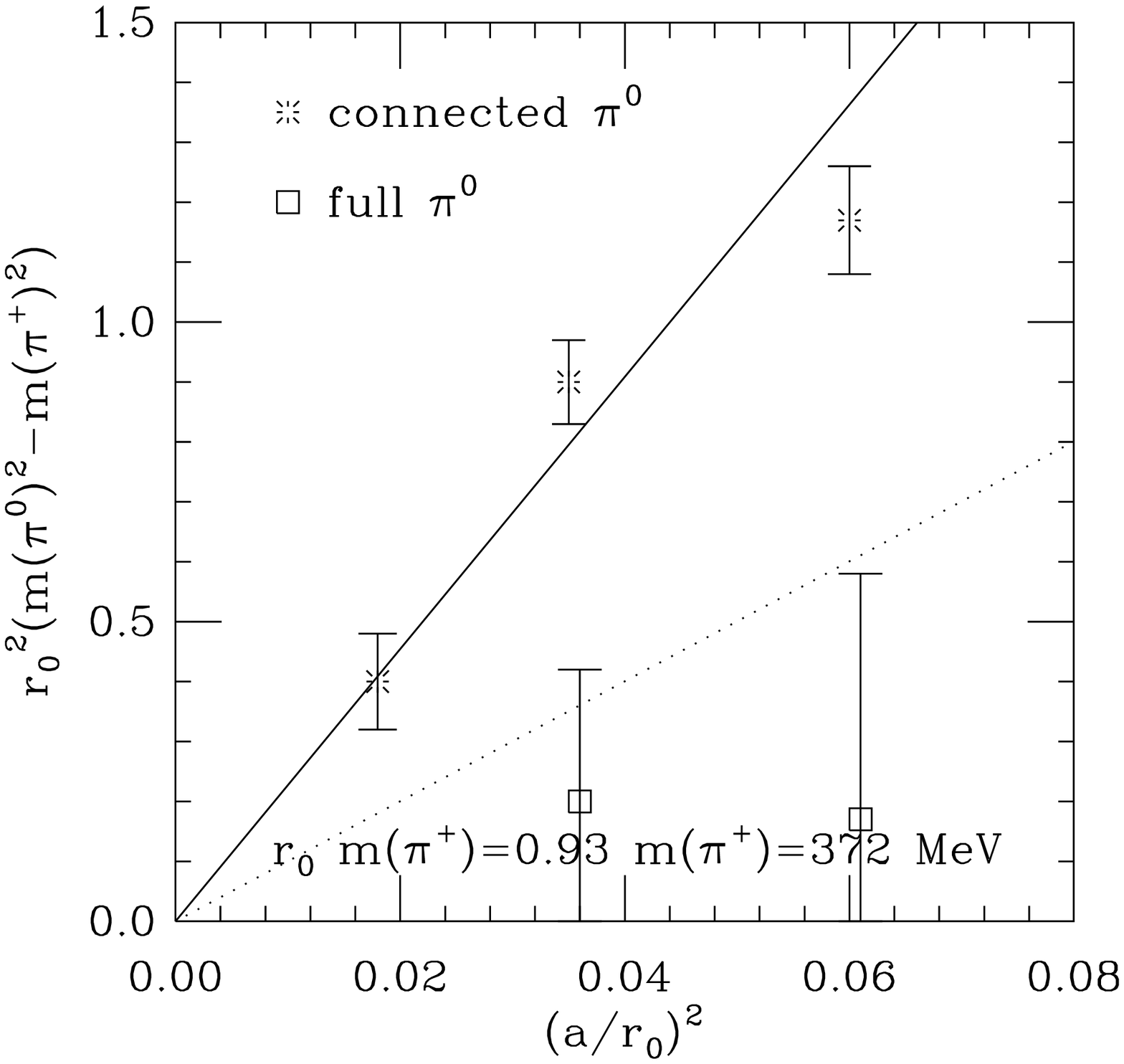,angle=0,width=0.48\linewidth}
\caption{Scaling behaviour of the mass splittings 
between the neutral and the charged pseudoscalar masses for 2 values of
$r_0m_{\rm PS}$. The open squared are the data for the neutral pseudoscalar
meson with tmQCD, and the stars only the connected contribution (pseudoscalar
meson with OS action). The full and dotted lines are an estimate of the $a^2$
dependence for the two pion splittings, making the hypothesis that O($a^2$)
effects are mass independent.}
\label{fig:pi0all}
\end{center}
\end{figure}

In fig. \ref{fig:pi0all} the scaling behaviour of the connected correlator (OS
pseudoscalar), is compared with the pseudoscalar meson in tmQCD where also the
disconnected part is included (in both computations method {\bf A} is used for
the determination of the critical mass).
\vspace{-0.5cm}
For all the technical details of the
computation of the disconnected part I refer to
\cite{Jansen:2005cg,Farchioni:2005hf}. The results show an O($a^2$) scaling
for both the pseudoscalar masses, even if there are indications that 
the neutral pseudoscalar meson for tmQCD (with the 
inclusion of the disconnected correlator) has 
reduced cutoff effects, within the rather large statistical errors. 
It is possible to give a very rough estimate of the pion splitting
$r_0^2(m_{\pi^0}^2 - m_{\pi^{\pm}}^2) \simeq c(a/r_0)^2$ with $c \simeq 10$
(with large errors).
Comparing to a quenched simulation for na\"ive staggered fermions with Wilson
gauge action \cite{Ishizuka:1993mt}, one finds a similar size of the flavour
splitting encountered for the pion mass at a similar lattice spacing with a
value $c \simeq 40$. For dynamical improved staggered fermions
a value of $c \simeq 10$ has been found \cite{Aubin:2004wf}.

\begin{figure}[htb]
\begin{center}
\epsfig{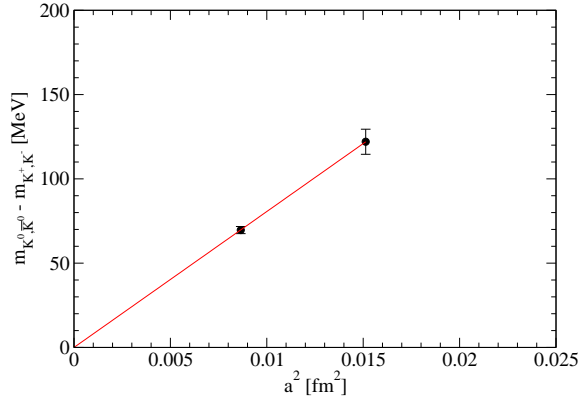}
\caption{Scaling behaviour of the mass splitting between neutral and charged kaons.}
\label{fig:kaons}
\end{center}
\end{figure}

An interesting study of the flavour breaking effects was presented at this
conference in \cite{Abdel-Rehim:2005qv}. To avoid the computation of
disconnected diagrams in the quenched approximation a second doublet for {\it
  strange} and {\it charm} quarks is introduced following the strategy of
\cite{Pena:2004gb}.
Then the splitting on the kaon system is studied. In this study method {\bf B}
has been used for the determination of the critical mass.
The results shown in fig. \ref{fig:kaons}, indicates that, as expected, the
flavour breaking effects vanish linearly with a rate of $a^2$, but that indeed
they could be significant at a lattice spacing $a > 0.1 $fm.
Recent results \cite{Farchioni:2005hf} with $N_{\rm f}=2$ dynamical tmQCD fermions and DBW2 gauge
action (see next sections for details on the simulation parameters), indicate
that at a lattice spacing of $a\simeq 0.12$ fm and a mass $\mu \simeq 12 $
MeV, the pion mass splitting even if with large errors, is consistent with
zero. 

\section{$N_{\rm f} =2$}

\subsection{Algorithmic improvements}

\begin{figure}[htb]
\begin{center}
\epsfig{file=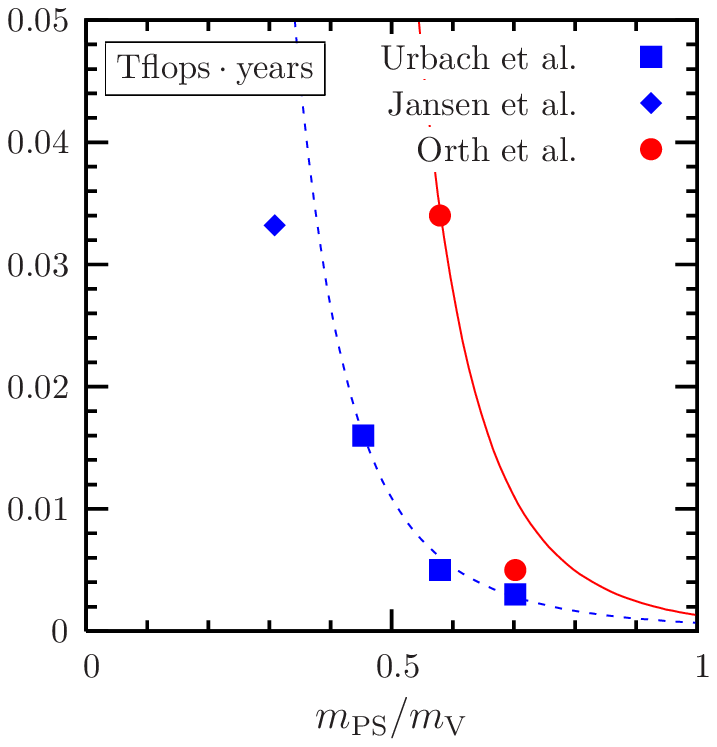,angle=0,width=0.4\linewidth}
\epsfig{file=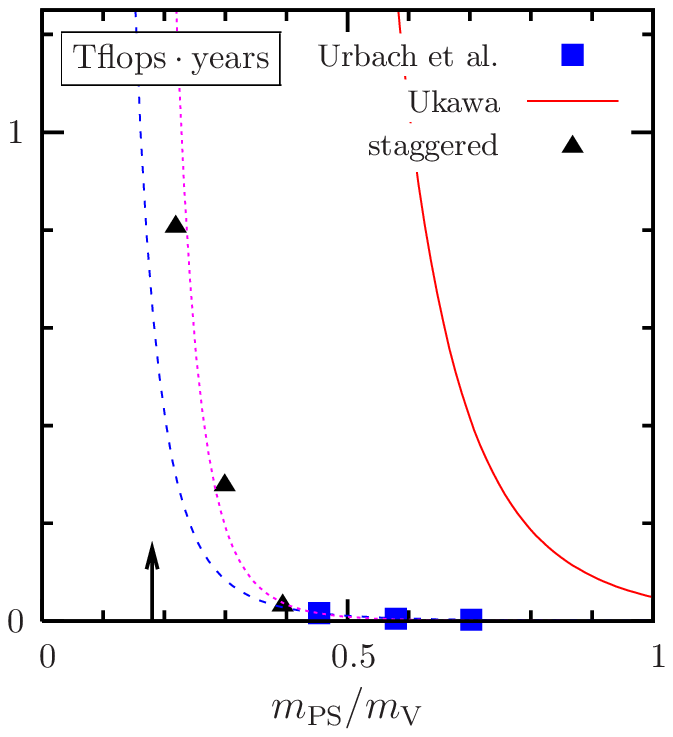,angle=0,width=0.4\linewidth}
\caption{Computer resources needed to generate $1000$ independent
    configurations of size $24^3\times 40$ at a lattice spacing of about
    $0.08\ \mathrm{fm}$ in units of $\mathrm{Tflops}\cdot
    \mathrm{years}$ as a function of $m_\mathrm{PS}/m_\mathrm{V}$. See text
    for a detailed description of the plots.}
\label{fig:bw2}
\end{center}
\end{figure}

In the Lattice 2001 conference A. Ukawa presented \cite{Ukawa:2002pc} a rather impressing
analysis on the possibility of simulating light quark masses with
Wilson fermions. This was summarized with the now well known Berlin wall
figure (see \cite{Jansen:2003nt} for a recent update).
Recently new algorithms \cite{Luscher:2004rx,Urbach:2005ji} 
have been proposed that have finally moved the wall to
rather small quark masses.
Both the algorithms are based on the standard HMC but have used new
preconditioner.
In \cite{Luscher:2004rx} it was shown that with a domain
decomposition (DD) preconditioning combined with a multiple time (mt) scale
integrator \cite{Sexton:1992nu}, light quark masses 
($m_\pi = 294$ MeV) are reachable with Wilson fermions with remarkable
performances.
In \cite{Urbach:2005ji} another very efficient preconditioner for the HMC algorithm has been
introduced and tested,
based on a mass preconditioner \cite{Hasenbusch:2001ne} (also known as Hasenbusch (H) acceleration) 
with again a multiple time scale integrator.
In table \ref{tab:algo} is summarized a rough comparison between the two algorithms, using
different lattice actions, based on the so called cost figure $\nu = 10^{-3}
(2N + 3)\tau_{\rm int}(P)$ introduced in \cite{Luscher:2004rx}. 
\begin{table}[b]
\begin{center}
\begin{tabular} {|c|c|c|c|c|c|}
\hline\hline
&&&&&\\[-0.5ex]
    Action &  Algorithms & $r_0/a$ & $m_{\rm \pi}$ [MeV] & $\nu$ & $\tau_{\rm int}$ \\  [1ex]
\hline
    W+W       & (mt)(DD)HMC \cite{Luscher:2004rx} & 6.40(15) & 294 & $0.74(18)$ & $21(5)$ \\
    tlSym+Wtm & (mt)(H)HMC  \cite{Urbach:2005ji}  & 5.20(25) & 280 & $0.49(34)$ & $21(14)$ \\
\hline\hline
\end{tabular}
\caption{Comparison of the 2 algorithms discussed in the text for a similar
  physical situation.}
\label{tab:algo}
\end{center}
\end{table}
The conclusion is that the algorithms have comparable performance down to pion
masses of the order of $m_\pi \simeq 300$ MeV.
In fig. \ref{fig:bw2} is plotted the update of the Berlin wall figure.
On the left panel there is a comparison between the results of
\cite{Urbach:2005ji,Jansen:2005yp} (squares and diamond) and the results of
\cite{Orth:2005kq} (circles). The lines are
functions proportional to $(m_\mathrm{PS}/m_\mathrm{V})^4$ (dashed) and
$(m_\mathrm{PS}/m_\mathrm{V})^6$ (solid).
On the right panel it is shown a comparison between the Ukawa's formula 
in \cite{Ukawa:2002pc} (solid line) and the extrapolation of the results in 
\cite{Urbach:2005ji} using a 
$(m_\mathrm{PS}/m_\mathrm{V})^4$ (dashed) and a
$(m_\mathrm{PS}/m_\mathrm{V})^6$ (dotted) dependence for the data. The arrow
indicates the physical pion to rho meson mass ratio. 
In addition there are also data points from staggered simulations 
(see \cite{Jansen:2003nt} and references therein). In particular this plot
indicates that running for one year a 1 Tflop sustained performance 
machine allows to generate at the physical point with $a \simeq 0.08$ 
fm and a lattice of $24^3 \times 40$,
1000 independent trajectories.

\subsection{Phase diagram of Wilson fermions}
\label{sec:phase}

In \cite{Farchioni:2004us} the first study of tmQCD with $N_{\rm f} = 2$
dynamical fermions was performed. Starting the exploration of a
completely new territory, it is always good to remember a sentence of
G. Parisi \cite{Parisi:1988hz} ``Let me describe a typical computer
simulation: the first thing to do is to look for phase transitions''.
It is important to have then the correct understanding of the phase diagram
with Wilson fermions in the 3 parameters space $(\beta = 6/g_0^2,m_0,\mu)$.
To check that the results are not induced by the algorithm used it is always good
to have at least 2 algorithms that reproduce the same results.

\begin{figure}[htb]
\epsfig{file=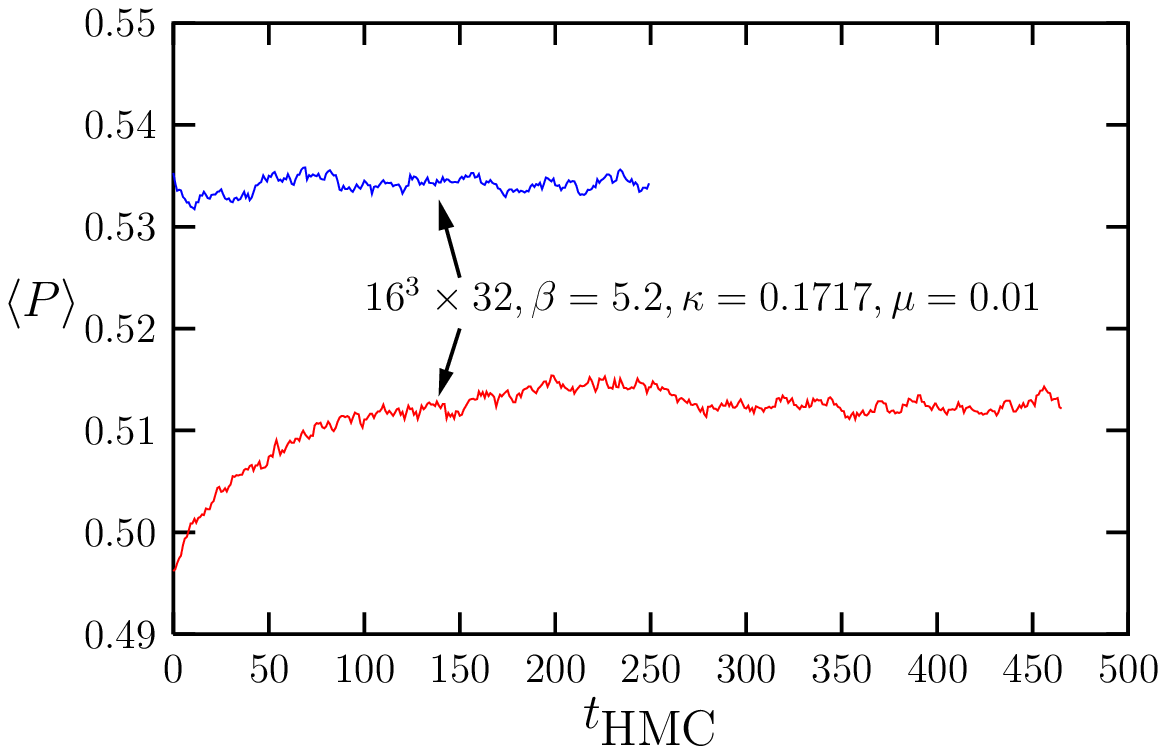,angle=0,width=0.5\linewidth}
\epsfig{file=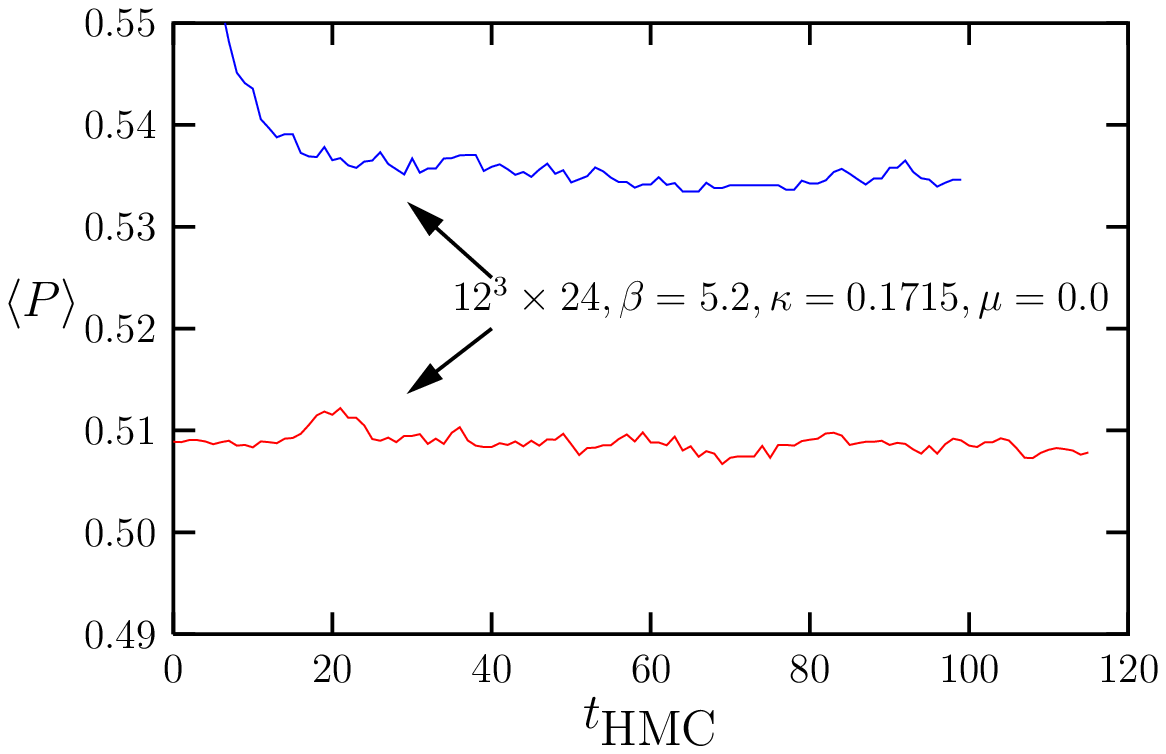,angle=0,width=0.5\linewidth}
\caption{Metastable states at $\beta=5.2$. Left panel: MC history of the
  average plaquette value with a twisted quark mass $\mu \simeq 10 $
  MeV and a lattice size $16^3 \times 32$. Right panel: MC history of the
  average plaquette for pure Wilson fermions ($\mu =0$)
  and a lattice size $12^3 \times 24$. }
\label{fig:meta}
\end{figure}

\begin{figure}[htb]
\epsfig{file=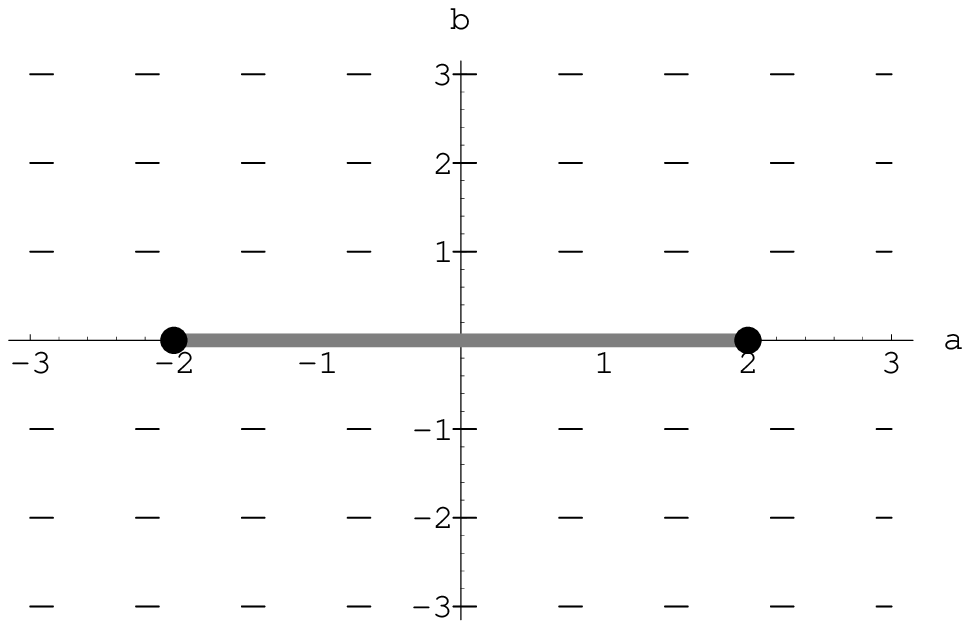,angle=0,width=0.5\linewidth}
\epsfig{file=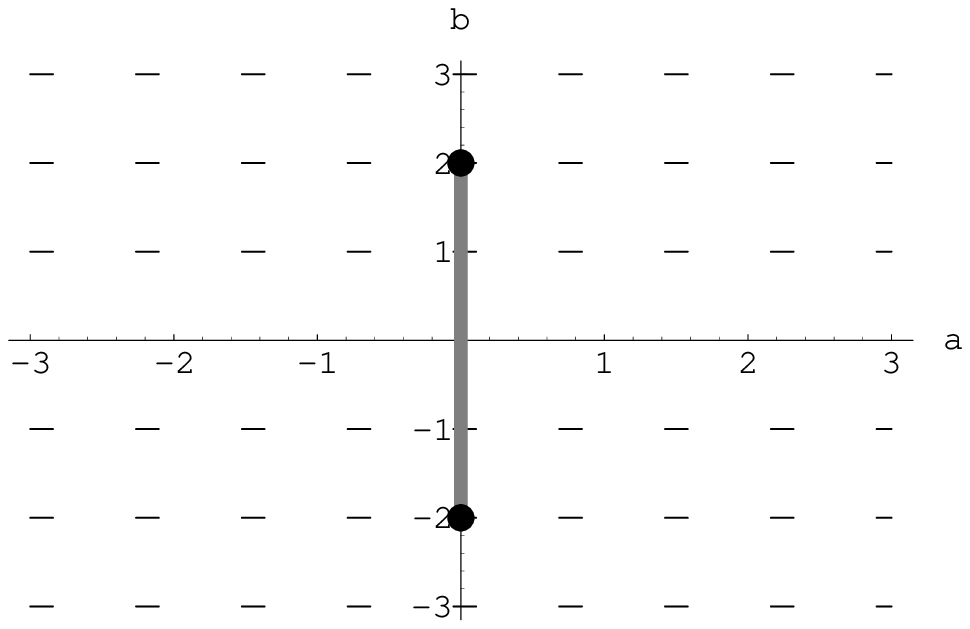,angle=0,width=0.5\linewidth}
\caption{Left panel: the phase diagram of Wilson diagram according to $\chi$
  PT for $c_2>0$. Right panel: as the left panel but for $c_2<0$. The x-axis
  is $m'/a^2$ and the y-axis is $\mu/a^2$.}
\label{fig:tmphase}
\end{figure}

Indeed in \cite{Farchioni:2004us} using the so called TSMB
\cite{Montvay:1995ea} and GHMC \cite{Hasenbusch:2001ne,Hasenbusch:2002ai}
algorithms rather surprising results were found. The action used was Wilson
gauge action combined with Wilson fermions with and without twisted mass.
In particular at a lattice spacing of $a\approx 0.16$ fm, strong evidence of a first
order phase transition was found for a rather large range of values of twisted
masses going from zero twisted mass to $\mu \simeq 100$ MeV. This study
reveals also that the phase transition tends to disappear increasing the value
of $\mu$, it persists for $\mu=0$ and it is volume independent.
A typical example of a MC history for the plaquette expectation value can be
seen in fig. \ref{fig:meta}, where a cold and a hot start was performed.
These results can help to see from a different point of view old numerical and
theoretical works. In \cite{Aoki:1995yf,Aoki:1996pw} from a finite
temperature study there was an indication of difficulties in observing a phase
with spontaneous breaking of flavour and parity symmetry (Aoki phase) at
$\beta > 4.8$.
In \cite{Blum:1994eh} the MILC collaboration found a surprising bulk first
order phase transition for Wilson fermions at $\beta \simeq 4.8$.
In \cite{Creutz:1996bg} an analysis using the linear sigma-model is
performed, finding an indication of two possible patterns of symmetry
breaking at finite lattice spacing. 
This observation was put on firmer theoretical basis in \cite{Sharpe:1998xm}.
In this very important paper several interesting results and consideration
were done, that, seen now from a different point a view, can help to understand
the rather surprising numerical results obtained in \cite{Farchioni:2004us}.
In \cite{Sharpe:1998xm} for the first time the concept of chiral lagrangian at
finite lattice spacing is given. The key point of the construction is the
observation that {\it the Pauli term in the effective Symanzik lagrangian
  transforms under chiral rotation exactly as does the mass term}. I would like
to add that this is also the key point for the automatic O($a$) improvement
for tmQCD at maximal twist.
Neglecting the derivative interaction, being interested in the vacuum state,
the potential of the effective chiral lagrangian reads
\be
{\mathcal V} = -{c_1 \over 4} \langle \Sigma+\Sigma^\dagger \rangle +
{c_2 \over 16} \langle \Sigma + \Sigma^\dagger \rangle^2
\ee
\be
c_1 \sim m'\Lambda^3 \qquad c_2 \sim m'^2\Lambda^2 + m'a\Lambda^4 +
      a^2\Lambda^6 \qquad m' = m-a\Lambda^2
\ee
where $\Sigma$ is the matrix that collects the Goldstone boson fields of the
theory. We remark here that $m'$ is a redefinition of the untwisted quark mass
that includes the O($a$) coming from the clover term. 
Up to O($a^2$) $m'$ is proportional to the PCAC quark mass.
The two terms in the potential become comparable when 
$m' \sim a^2\Lambda^3$. In this region of quark masses the competition of
these two terms causes a non-trivial vacuum structure that gives the following 
2 scenarios \cite{Sharpe:1998xm} for the phase diagram of Wilson fermions: 1) The Aoki phase
\cite{Aoki:1984qi}; 2) The existence of a $1^{\rm st}$ order phase transition 
\cite{Sharpe:1998xm}.
The extension to tmQCD of these results is done in
\cite{Munster:2004am,Scorzato:2004da,Sharpe:2004ps}. The result is 
summarized in fig. \ref{fig:tmphase} where the x-axis is $m'/a^2$ and the
y-axis is $\mu/a^2$. A non-zero value of the twisted mass washes out the Aoki
phase introducing an explicit breaking of flavour and parity symmetry. As can
be seen from fig. \ref{fig:tmphase} (left panel) the Aoki phase lies on the untwisted
axis. In the second scenario in fig. \ref{fig:tmphase} (right panel) 
the first order phase transition line extends into
the twisted direction to a distance of $\mu_{\rm c} \approx a^2\Lambda^3$. The transition ends
with a second order phase transition point, where the neutral pion mass vanishes.
Several comments are in order now.
The occurrence of one of the two scenarios depends on the sign of the coefficient
$c_2$ proportional to the O($a^2$) term in the chiral lagrangian.
This coefficient $c_2$ depends on the choice of the gauge action, on the
presence in the lattice action of the clover term and on the bare gauge coupling.
An analysis with Wilson fermions of the two dimensional Gross-Neveau model
\cite{Izubuchi:1998hy} indicates that indeed both the scenarios describe
the phase structure of Wilson fermions depending on the value of the 
couplings of the model. The analysis shows that at strong coupling there is an Aoki phase
while at weak coupling the first order phase transition line sets in.
This analysis has been recently extended for the twisted mass case
\cite{Nagai:2005mi}, indicating even more complicated structures, like a coexistence
of the two scenarios at the same value of the coupling.

\begin{figure}[htb]
\vspace{-0.0cm}
\begin{center}
\epsfig{file=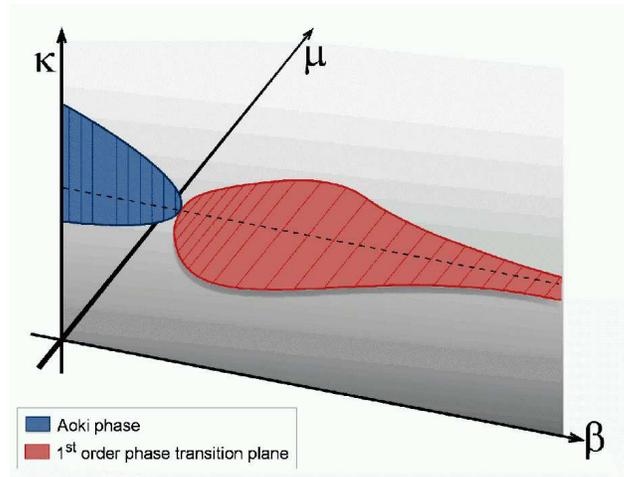,angle=0,width=0.55\linewidth}
\caption{Current knowledge of the Wilson lattice QCD phase diagram as function of the inverse
gauge coupling $\beta\propto 1/g^2$, the hopping parameter $\kappa$ and the
twisted mass parameter $\mu$.}
\label{fig:phase}
\end{center}
\end{figure}

Our present understanding of the lattice QCD phase diagram can be 
summarized as following.
For values of the lattice spacing much coarser than $a=0.15\
\mathrm{fm}$ there is a second order phase transition from the standard 
lattice QCD phase to
the Aoki phase \cite{Aoki:1984qi,Ilgenfritz:2003gw,Sternbeck:2003gy}. 
For smaller values of the lattice spacing a first order phase transition
appears \cite{Farchioni:2004us,Farchioni:2004fs,Farchioni:2004ma,Farchioni:2005tu}
that separates the positive quark mass from the negative quark mass
phase. This first order phase transition is reminiscent of the continuum 
phase transition when the quark mass is changed from positive to negative values
with the corresponding jump of the scalar condensate as the order parameter of
spontaneous chiral symmetry breaking. 
The generic phase structure of lattice QCD 
is illustrated in fig.~\ref{fig:phase} and discussed in
refs.~\cite{Farchioni:2004us,Farchioni:2004fs,Farchioni:2004ma}.

\subsection{Minimal pion mass}

In the scenario with a first order phase transition the pseudoscalar mass
$m_{\rm PS}$ cannot be made arbitrarily small both if the chiral point is reached
from the untwisted or the twisted direction. Lowering the quark mass from 
the untwisted direction the
algorithm will start to sample also in the region with negative masses.
The minimal pion mass reachable will then depend on the algorithm used and on
the strength of the phase transition.
Lowering the quark mass from the twisted direction there is a minimal pion mass given directly by the
extension of the first order phase transition line, even if the twisted mass
gives a sharp infrared cutoff in the sampling performed by the algorithm.

It therefore becomes important to 
understand the phase structure of lattice QCD
as a pre-requisite before starting large scale
simulations. As we have seen the extension of the first order phase transition
line in the twisted direction is proportional to the coefficient $|c_2|$.
This coefficient depends both on the gauge action used and on the presence of
the clover term in the lattice action.
In \cite{Farchioni:2005tu} has been studied the lattice spacing dependence of
of the first order phase transition with Wilson gauge action, 
taking as a measure of its strength, the
gap  between the two phases in the plaquette expectation value and in the PCAC
quark mass.
The qualitative estimate for the lattice spacing, where a minimal pion mass
$m_\pi \simeq 300 $ MeV could be reached, without being affected by the first
order first transition is 0.07-0.1 fm.

It is suggestive that at the microscopic level the occurrence of this first
order phase transition is accompanied by a massive rearrangement 
of the small eigenvalues of the Wilson-Dirac. This rearrangement could be
suppressed by the use of a renormalization group improved or O($a^2$) improved
gauge actions, and indeed results from \cite{Aoki:2004iq} indicate that
metastabilities in the average plaquette observed for 
$N_{\rm f} = 3$ dynamical Wilson fermions with a clover term (there is also an
indication that the same metastabilities survive without a clover term for $N_{\rm f} = 3$),
can be suppressed replacing the Wilson gauge action with the Iwasaki
action \cite{Iwasaki:1985we}. 

\subsection{Tree-level Symanzik improved gauge action}

The dependence of the phase diagram on the gauge action used and on the
lattice spacing has been studied in a set of papers 
\cite{Farchioni:2004us,Farchioni:2004fs,Farchioni:2004ma,Farchioni:2005tu}
(see also \cite{Farchioni:2005ec} for a detailed summary of these results).
The gauge actions so far studied can be parameterized by
\be
    S_{\rm G} = \beta \big[b_0\sum_{x;\mu<\nu}(1-{1 \over 3}P^{1 \times
    1}(x;\mu,\nu)) + b_1 (1-{1 \over 3} P^{1 \times 2}(x;\mu,\nu)) \big]
\ee
with the normalization condition $b_0 = 1-8b_1$.
The parameters of the tree-level Symanzik action \cite{Weisz:1982zw} 
($b_1 = - {1 \over 12}$) simulations are summarized in tab. \ref{tab:tlsym}.
The last line indicates an estimate of the minimal pion mass reachable at the
corresponding lattice spacings.
\begin{table}
\begin{center}

  \begin{tabular}{|l|l|l|}\hline

     \multicolumn{1}{|c|}{ $\beta=3.65$  } & 
     \multicolumn{1}{|c|}{ $\beta=3.75$  } & 
     \multicolumn{1}{|c|}{ $\beta=3.90$  } 
    \\ 
\hline
\ $a\mu = 0.01$                        & \   $a\mu =0.0094-0.005$             & \   $a\mu =0.0075-0.004$     \\
\ $a\approx 0.13$ fm                   & \   $a\approx 0.12 $fm               & \   $a\approx 0.1$ fm    \\
\ $L\approx 1.56$ fm                   & \   $L\approx 2 $fm                  & \   $L\approx 1.6$ fm   \\
\ $(m_\pi)_{\rm min} \approx 450$ MeV  & \   $(m_\pi)_{\rm min} \approx 400$
     MeV   & \   ($m_\pi)_{\rm min} \approx 280$ MeV  \\
\hline
\end{tabular}
\caption{Summary of the simulation parameters for dynamical runs of tmQCD with
  tlSym gauge action. The last line is an estimate of the minimal pion mass
  reachable without encountering metastabilities.}
\label{tab:tlsym}
\end{center}
\end{table}
In order to check for a possible phase transition and corresponding
metastabilities a measure of the average plaquette value as a function of the
hopping parameter $\kappa$ on runs that start from both a hot and a cold
configuration has to be done. 
Since the metastability, if any, will show up around $\kappa_c$ (determined
monitoring the PCAC mass $m_{\rm PCAC}$ at the corresponding fixed value of $\mu$)
attention should be given to the hot and cold runs on $\kappa$-values closest to $\kappa_c$
only.

\begin{figure}[htb]
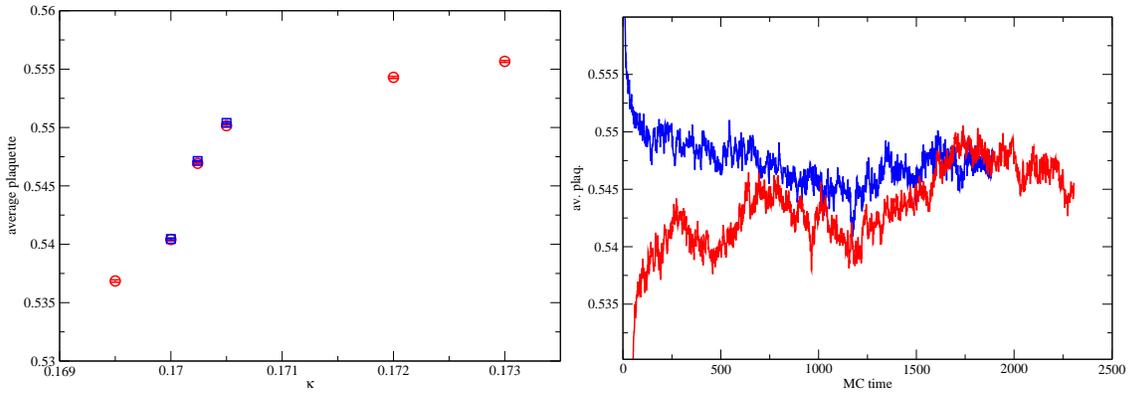

\begin{center}
\includegraphics[width=0.49\linewidth]{plots/av_plaq_tlSym_b3.65_m0.01_L12T24.eps}
\includegraphics[width=0.49\linewidth]{plots/L12T24_b3.65_m0.01_plaq_mc_history.eps}
\caption{Left panel: Average plaquette value 
vs.~$\kappa$ at $\beta=3.65, r_0 \mu=0.038$
  on a $12^3 \times 24$ lattice from hot (red symbols) and cold starts (blue symbols).
Right panel: 
Average plaquette MC time history for two runs at $\beta=3.65, r_0 \mu=0.038, \kappa=0.17024$
  on a $12^3 \times 24$ lattice starting from hot (red line) and cold configuration
  (blue line).}
\label{fig:b3.65}
\end{center}
\end{figure}

At $\beta=3.65, a \simeq  0.13 fm, 12^3 \times 24$ and
$\mu \simeq 15$ MeV there are signs of a very nearby phase transition, as can be deduced
from the steep rise in $\kappa$ of the plaquette expectation value (left panel
in fig. \ref{fig:b3.65}), from a very slow
thermalization and large fluctuations of the plaquette MC history value over several
hundreds of trajectories (right panel in fig. \ref{fig:b3.65}).
An estimate of the pseudoscalar mass, close to $\kappa_{\rm c}$, is $m_{\rm PS} \simeq 450 MeV$.

\begin{figure}[htb]
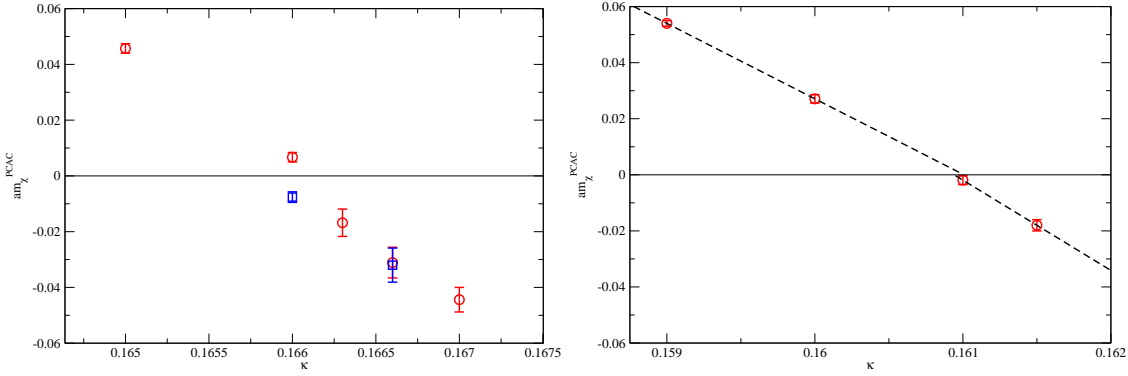

\vspace{-0.0cm}
\begin{center}
\includegraphics[width=0.49\linewidth]{plots/tlSym_L12T24_b3.75_m0.005_mPCAC_vs_kappa.eps}
\includegraphics[width=0.49\linewidth]{plots/tlSym_L16T32_b3.90_m0.0075_mPCAC_vs_kappa.eps}
\caption{PCAC quark mass $m_{\rm PCAC}$ vs.~$\kappa$ on a $16^3 \times 32$
  lattice. Left panel: $a \simeq
  0.12 $ fm, $\mu \simeq 8 $ MeV. Right panel: $a \simeq 0.1$ fm, $\mu \simeq
  15$ MeV.}
\label{fig:b3.75}
\end{center}
\end{figure}

At $\beta=3.75, a \simeq  0.12$ fm, $12^3 \times 24$ and 
$\mu \simeq 8$ MeV there is a similar situation we have observed before at
$\beta=3.65$ and $\mu \simeq 15$ MeV.
This is described by fig. \ref{fig:b3.75} (left panel), where it
is plotted the $\kappa$ dependence of the PCAC mass.
This dependence is very useful to monitor a possible metastable critical
point, since this shows up in a different extrapolated $\kappa_{\rm c}$, when
the extrapolation is performed from positive or negative quark masses.
A second twisted mass $\mu \simeq 15$ MeV has been simulated in a lattice $16^3
\times 32$ around the
critical point for this lattice spacing. Even if a strict check done with a
hot and a cold start is not available at the moment the $\kappa$ dependence of
the PCAC mass for this second value of $\mu$ suggests that the critical point
is free from metastabilities. The pseudoscalar mass measured for the heaviest twisted
mass is around $m_{\rm PS} \simeq 400$ MeV.
At $\beta=3.9, a \simeq  0.1$ fm, $16^3 \times 32$ and $\mu \simeq 8$ and $15$
MeV, there are no signs of metastabilities at the two corresponding critical
points.
In fig. \ref{fig:b3.75} (right panel) is plotted the $\kappa$ dependence
of the PCAC quark mass.
The pseudoscalar masses obtained for the two values of $\mu$ are respectively
$m_{\rm PS} \simeq 280$ and $450$ MeV.
We remark also that the physical volume at this $\beta$ value is rather small
$L\simeq1.6$ fm, and results obtained for pure Wilson fermions
\cite{Luscher:2004rx,Orth:2005kq}, indicate that for these quark masses and
these volumes the finite size effects could be substantial. The estimate of
the minimal pseudoscalar mass for this lattice spacing is then clearly only 
an upper bound.

\subsection{DBW2 gauge action}

\begin{figure}[htb]
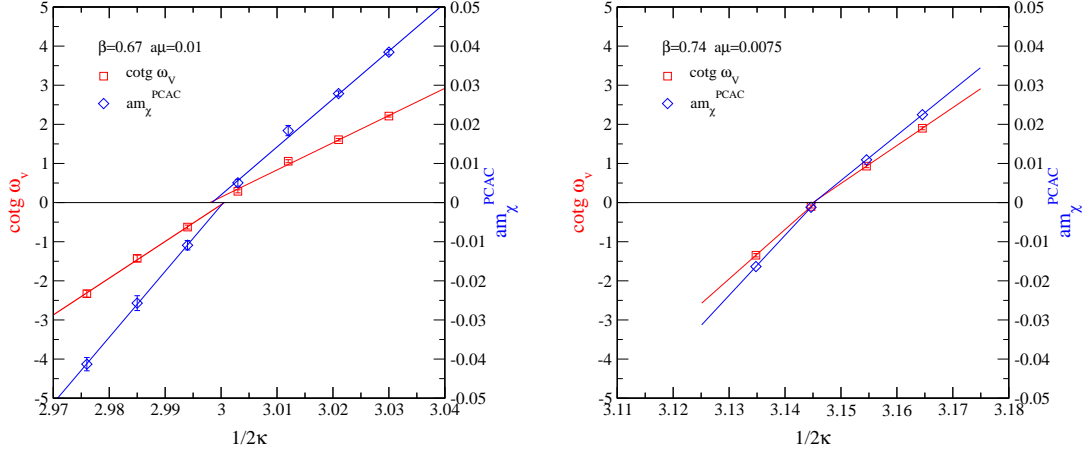

\vspace{+1cm}
\begin{center}
\epsfig{file=plots/full_tw_extra_b067.eps,angle=0,width=0.45\linewidth}
\hspace{0.5cm}
\epsfig{file=plots/full_tw_extra_b074.eps,angle=0,width=0.45\linewidth}
\caption{Determination of the critical hopping parameter 
$\kappa_{\rm c}$ by extrapolating to zero the untwisted PCAC quark mass $m_{\rm PCAC}$.
The small discrepancy observed at $\beta=0.67$ (left panel) 
between extrapolations from positive and negative
quark masses is probably a small effect of the first order phase transition.
For $\beta=0.74$ (right panel), extrapolations
from both sides give consistent results. An alternative way to fix the
critical mass is also plotted (see \cite{Farchioni:2005ec} for details).}
\label{fig:dbw2}
\end{center}
\end{figure}

In this section I summarize the results \cite{Farchioni:2004fs} 
obtained using the so called DBW2 gauge action \cite{Takaishi:1996xj} ($b_1 = -1.4088$).
The parameters used in the simulations are summarized in tab. \ref{tab:dbw2}.
The twisted mass for the two lattice spacing is kept roughly fixed to $\mu
\approx 12$ MeV.
The last line indicates an estimate of the minimal pion mass reachable at the
corresponding lattice spacings.
\begin{table}
\begin{center}

  \begin{tabular}{|l|l|}\hline

     \multicolumn{1}{|c|}{\ $\beta=0.67$ } & 
     \multicolumn{1}{|c|}{\ $\beta=0.74$ } 
    \\ 
\hline
\ $ a\mu = 0.01$                & \   $a\mu = 0.0075$                    \\
\ $a\approx 0.19$ fm                & \   $a\approx 0.12$ fm                    \\
\ $L\approx 2.3$ fm                 & \   $L\approx 2 $ fm                       \\
\ $(m_\pi)_{\rm min}\approx 360$ MeV    & \   $(m_\pi)_{\rm min}\approx 320$ MeV        \\
\hline
\end{tabular}
\caption{Summary of the simulation parameters for dynamical runs of tmQCD with
  DBW2 gauge action. The last line is an estimate of the minimal pion mass
  reachable without encountering metastabilities.}
\label{tab:dbw2}
\end{center}
\end{table}
Also for this gauge action several quantities have been computed. Here we
concentrate as before on the PCAC mass and on the minimal pion mass.
In contrast with the tlSym results here simulations at full twist
were never performed, so the evidence for a metastability region can be
deduced only indirectly from the dependence of the PCAC mass on the untwisted
quark mass as discussed before.
In fig. \ref{fig:dbw2} is shown the $1/(2\kappa)$ dependence of the PCAC mass
for the two lattice spacing used. At $a \approx 0.19$ fm there is an indirect
evidence of a small metastability at full twist, that seems to disappear at
$a \approx 0.12$ fm.
To summarize, there are first indications, that in order to reach pion masses
of the order of $m_{\rm PS} \simeq 300$ MeV with tmQCD a gauge action like
tlSym or DBW2 is appropriate. There are also first indications that this pion
mass can be reached with DBW2 at slightly coarser lattices. In order to avoid
possible large cutoff effects with DBW2 (see for example fig. 3 in
\cite{Necco:2003vh}), or big coefficients in perturbative expansions, with the
present data, the tlSym gauge action may be considered a better choice.

\subsection{$N_{\rm f} = 2+1+1$}

The fact that tmQCD can be formulated only for an even number of flavours is not a
limitation. Indeed using an off diagonal splitting, where the degenerate quark
doublet has a different flavour orientation from the splitting between the quarks, in
\cite{Frezzotti:2003xj} it was shown that the determinant for non degenerate
quarks is real and positive (see \cite{Pena:2004gb} for alternative
formulations of tmQCD including a non degenerate doublet).
The only restriction of the construction in \cite{Frezzotti:2003xj} 
is on the value of the ratio between renormalization constants of the
pseudoscalar and scalar current. To give an example, fixing the values of the
renormalized {\it strange} and {\it charm} quark masses, gives the following constraints
\be
 \mu^{\rm R}_{\rm c} \simeq 1.5 {\rm GeV} \qquad  \mu^{\rm R}_{\rm s} \simeq 0.1 {\rm GeV}
  \Rightarrow {Z_{\rm P} \over Z_{\rm S}} > 0.875.
\ee
At this conference first results with dynamical $N_{\rm f} = 2+1+1$ twisted
quarks have been presented \cite{Farchioni:2005ec}.
The simulations of the $N_{\rm f} = 2+1+1$ theory are performed by a
polynomial hybrid Monte Carlo algorithm (PHMC)\cite{Frezzotti:1997ym}. 
The structure of the algorithm goes along the lines indicated in
\cite{Montvay:2005tj}. At this conference another variant of the PHMC to
include the two non degenerate twisted quarks has been presented \cite{Chiarappa:2005mx}.

\section{Further results}

In this section I summarize further results concerning tmQCD. 
In \cite{Frezzotti:2005my} it has been presented a strategy 
to compute $B_{\rm K}$ and matrix elements related 
to the $\Delta I = 1/2$ rule without mixing with operators with wrong
chiralities, retaining all the properties of automatic O($a$) improvement.
The strategy is based on the usage of a mixed action (OS for valence quarks
and tmQCD for sea quarks) \cite{Frezzotti:2004wz}.
In the quenched approximation $B_{\rm K}$ has been
computed \cite{Dimopoulos:2004xc} in the continuum limit, using clover improved tmQCD and
a non-perturbative renormalization without mixing (in the SF scheme).
A strategy to compute $B_{\rm B}$ with tmQCD, along the lines of \cite{Guagnelli:2002xz}, has
been proposed in \cite{Palombi:2005pa}, and along the lines of \cite{Frezzotti:2004wz} in
\cite{DellaMorte:2004wn}.
In \cite{Pena:2004gb} a strategy, based on clover improved tmQCD with $N_{\rm
  f} =4$, has been proposed to compute the renormalization of $K
\rightarrow \pi$ matrix elements.
In \cite{Gattringer:2005vp} the effect of a twisted mass term of the low-lying modes of the
  Wilson-Dirac operator and a remnant of the index theorem for twisted mass
  fermions has been discussed.

\section{Conclusions}

Several lessons come from quenched studies of tmQCD. With a particular and field
theoretically well founded definition of the critical mass, automatic O($a$)
improvement is effective till small pion masses ($m_\pi = 272$ MeV), and the
residual O($a^2$) cutoff effects are small. The bending phenomenon just results
from big cutoff effects, that are reproducible with $\chi$PT at finite lattice
spacing.
The bending phenomenon is not present even at 
finite lattice spacing with a suitable choice of the critical mass.
The flavour breaking is an issue and it has to be investigated with dynamical
simulations.
We have indications of the existence of an Aoki phase for quenched Wilson
fermions at lattice spacings around $a \simeq 0.1$ fm.

To perform dynamical simulations at small pion masses,
algorithmic improvements are crucial, and now new algorithms 
allow to have efficient and performant simulations with Wilson
fermions and most probably with staggered fermions.

We have a much better understanding of the phase structure of dynamical Wilson
fermions. A theoretically well founded action (tlSym gauge and tmQCD fermion
action) allows to perform dynamical simulations with $N_{\rm f}=2$ at pion masses
smaller then 300 MeV starting from a lattice spacing $a \simeq 0.1$ fm, 
allowing matching with $\chi$PT, and simulation with
$N_{\rm f}=2 +1+1$ flavours are just starting.

In many cases it has been shown that the renormalization properties of local
operators related to very important phenomenological quantities, is continuum
like.

Although presently not all aspects of tmQCD are fully investigated, tmQCD is
an attractive and powerful discretization of lattice QCD, and it
certainly belongs to the pool of well founded fermion actions that ought to be
used to control the continuum limit of physical quantities of interest.

\vspace{0.5cm}

{\bf Acknowledgments}:
I thank the LOC for the stimulating atmosphere of the conference.
I would like to thank F. Farchioni, I. Montvay, K. Nagai, 
M. Papinutto, E. Scholz, L. Scorzato, N. Ukita, 
C. Urbach, U. Wenger and I. Wetzorke for discussions and a most 
enjoyable collaboration, and in particular Karl Jansen for constant 
encouragement, and for a careful reading of this manuscript.
I have profited from interesting and useful discussions with
S. Aoki, O. B\"ar, R. Frezzotti, C. Michael, 
G. C. Rossi, S. Sharpe, S. Sint.
This work is partially supported by the DFG Sonderforschungbereich/Transregio
SFB/TR9-03.

\bibliographystyle{JHEP-2}
\bibliography{proc}

\end{document}